\documentclass[prl,aps,amsfonts,amssymb,amsmath,floats,floatfix,twocolumn,superscriptaddress,unsortedaddress]{revtex4}

\usepackage{epsfig,graphicx}
\usepackage{graphics}
\usepackage{float}
\usepackage{multirow}
\usepackage{amsmath}
\usepackage{times}
\usepackage{tabularx, booktabs}
\newcolumntype{Y}{>{\centering\arraybackslash}X}
\usepackage{array}
\newcolumntype{P}[1]{>{\centering\arraybackslash}p{#1}}

\begin{document}

\title{The symmetry of charge order in cuprates}

\author{R. Comin}
\email{r.comin@utoronto.ca}
\affiliation{Department of Physics {\rm {\&}} Astronomy, University of British Columbia, Vancouver, British Columbia V6T\,1Z1, Canada}
\affiliation{Quantum Matter Institute, University of British Columbia, Vancouver, British Columbia V6T\,1Z4, Canada}
\author{R. Sutarto}
\affiliation{Canadian Light Source, Saskatoon, Saskatchewan  S7N 2V3, Canada}
\author{F. He}
\affiliation{Canadian Light Source, Saskatoon, Saskatchewan  S7N 2V3, Canada}
\author{E. H. da Silva Neto}
\affiliation{Department of Physics {\rm {\&}} Astronomy, University of British Columbia, Vancouver, British Columbia V6T\,1Z1, Canada}
\affiliation{Quantum Matter Institute, University of British Columbia, Vancouver, British Columbia V6T\,1Z4, Canada}
\affiliation{Max Planck Institute for Solid State Research, Heisenbergstrasse 1, D-70569 Stuttgart, Germany}
\affiliation{Quantum Materials Program, Canadian Institute for Advanced Research, Toronto, Ontario M5G 1Z8, Canada}
\author{L. Chauviere}
\affiliation{Department of Physics {\rm {\&}} Astronomy, University of British Columbia, Vancouver, British Columbia V6T\,1Z1, Canada}
\affiliation{Quantum Matter Institute, University of British Columbia, Vancouver, British Columbia V6T\,1Z4, Canada}
\affiliation{Max Planck Institute for Solid State Research, Heisenbergstrasse 1, D-70569 Stuttgart, Germany}
\author{A.\,Frano}
\affiliation{Max Planck Institute for Solid State Research, Heisenbergstrasse 1, D-70569 Stuttgart, Germany}
\affiliation{\mbox{Helmholtz-Zentrum Berlin f\"{u}r Materialien und Energie, Wilhelm-Conrad-R\"{o}ntgen-Campus BESSY II, Berlin, Germany}}
\author{R. Liang}
\affiliation{Department of Physics {\rm {\&}} Astronomy, University of British Columbia, Vancouver, British Columbia V6T\,1Z1, Canada}
\affiliation{Quantum Matter Institute, University of British Columbia, Vancouver, British Columbia V6T\,1Z4, Canada}
\author{W.N. Hardy}
\affiliation{Department of Physics {\rm {\&}} Astronomy, University of British Columbia, Vancouver, British Columbia V6T\,1Z1, Canada}
\affiliation{Quantum Matter Institute, University of British Columbia, Vancouver, British Columbia V6T\,1Z4, Canada}
\author{D.A.\,Bonn}
\affiliation{Department of Physics {\rm {\&}} Astronomy, University of British Columbia, Vancouver, British Columbia V6T\,1Z1, Canada}
\affiliation{Quantum Matter Institute, University of British Columbia, Vancouver, British Columbia V6T\,1Z4, Canada}
\author{Y.\,Yoshida}
\affiliation{National Institute of Advanced Industrial Science and Technology (AIST), Tsukuba, 305-8568, Japan}
\author{H. Eisaki}
\affiliation{National Institute of Advanced Industrial Science and Technology (AIST), Tsukuba, 305-8568, Japan}
\author{A.J.\,Achkar}
\affiliation{Department of Physics and Astronomy, University of Waterloo, Waterloo, N2L 3G1, Canada}
\author{D.G.\,Hawthorn}
\affiliation{Department of Physics and Astronomy, University of Waterloo, Waterloo, N2L 3G1, Canada}
\author{B. Keimer}
\affiliation{Max Planck Institute for Solid State Research, Heisenbergstrasse 1, D-70569 Stuttgart, Germany}
\author{G.A. Sawatzky}
\affiliation{Department of Physics {\rm {\&}} Astronomy, University of British Columbia, Vancouver, British Columbia V6T\,1Z1, Canada}
\affiliation{Quantum Matter Institute, University of British Columbia, Vancouver, British Columbia V6T\,1Z4, Canada}
\author{A. Damascelli}
\email{damascelli@physics.ubc.ca}
\affiliation{Department of Physics {\rm {\&}} Astronomy, University of British Columbia, Vancouver, British Columbia V6T\,1Z1, Canada}
\affiliation{Quantum Matter Institute, University of British Columbia, Vancouver, British Columbia V6T\,1Z4, Canada}

\maketitle

{\bf Charge-ordered ground states permeate the phenomenology of \textit{3d}-based transition metal oxides, and more generally represent a distinctive hallmark of strongly-correlated states of matter. The recent discovery of charge order in various cuprate families fueled new interest into the role played by this incipient broken symmetry within the complex phase diagram of high-${T}_{\mathrm{c}}$ superconductors. Here we use resonant X-ray scattering to resolve the main characteristics of the charge-modulated state in two cuprate families: Bi${}_{2}$Sr${}_{2-x}$La${}_{x}$CuO${}_{6+\delta}$ (Bi2201) and YBa${}_{2}$Cu${}_{3}$O${}_{6+y}$ (YBCO). We detect no signatures of spatial modulations along the nodal direction in Bi2201, thus clarifying the \textit{inter-unit-cell} momentum-structure of charge order. We also resolve the \textit{intra-unit-cell} symmetry of the charge ordered state, which is revealed to be best represented by a bond-order with modulated charges on the O-2\textit{p} orbitals and a prominent \textit{d}-wave character. These results provide insights on the microscopic description of charge order in cuprates, and on its origin and interplay with superconductivity.}
\\

Complex oxides exhibit a mosaic of exotic electronic phases with various symmetry-broken ground states that revolve around three main instabilities: antiferromagnetism, charge order, and superconductivity. In particular, charge order -- the tendency of the valence electrons to segregate into periodically-modulated structures -- is found in various classes of strongly-correlated \textit{3d}-oxides, such as manganites \cite{Yoshizawa_1995}, nickelates \cite{Tranquada_1996}, and cobaltates \cite{Cwik_2009}. The original discovery of period-4 stripe-like charge correlations in the La-based materials \cite{tranquada1995,vZimmermann_1998,abbamonte2005,Fink2009} confirmed the central role played by charge-ordered states in the physics of underdoped cuprates, as anticipated by earlier theoretical work \cite{Poilblanc1989,Zaanen1989,Machida1989,Emery1990,Castellani1995}. Following further indications by surface-sensitive scanning tunnelling microscopy (STM) \cite{hoffman2002,howald2003}, the field was recently revived
by the detection of charge-modulated states in YBCO using nuclear magnetic resonance \cite{wu2011} and resonant X-ray scattering (RXS), with wavevector ${Q}^{*}\!\sim\!0.31$ reciprocal lattice units (r.l.u., used hereafter) \cite{ghiringhelli2012,chang2012,achkar2012,Blackburn2013,Blanco2013,LeTacon2013}.
\begin{figure}[h!]
\includegraphics[width=0.95\linewidth]{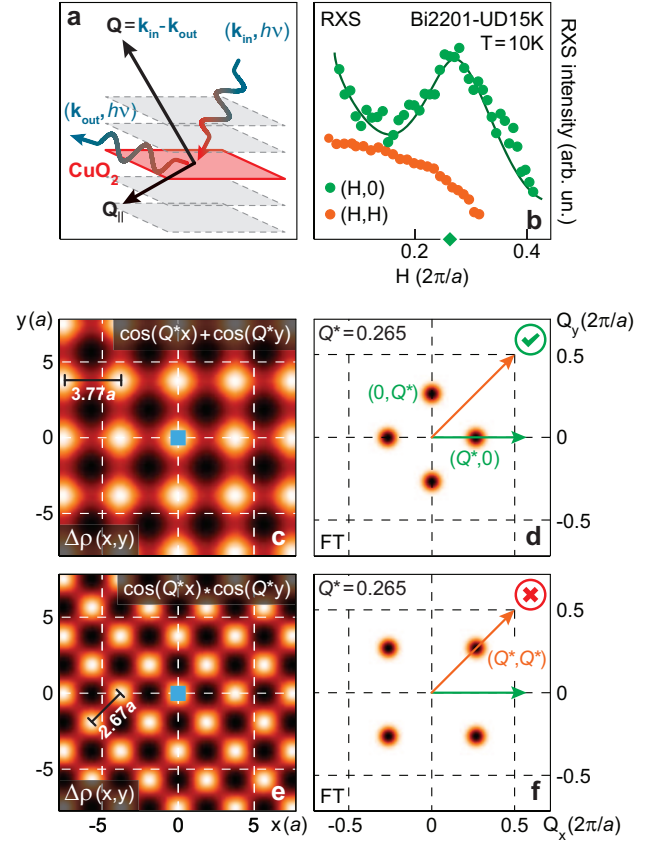}
\caption{{\bf Charge ordering patterns and wavevectors}. \textbf{a}, Schematics of a RXS experiment. \textbf{b}, Low-temperature RXS (at photon energy $h \nu\!=\!931.5$\,eV) from an underdoped Bi2201-UD15K sample, mapping reciprocal-space features along the two high-symmetry directions: $\left( H,0 \right)$, antinodal, green (reproduced from Ref.\,\onlinecite{Comin_Science}, the full line represents a Gaussian fit plus background); and $\left( H,H \right)$, nodal, orange. \textbf{c,e}, Modulation of the charge density $\Delta \rho (x,y)$, with functional form given by a sum (c) and product (e) of cosines, and a wavevector magnitude ${Q}^{*}\!=\!0.265$ r.l.u. (black bars indicate the period and direction of the spatial modulation, expressed in terms of the lattice parameter $a\!=\!3.86$\,\AA). The blue rectangles denote the undistorted unit cell. \textbf{d,f}, Fourier transforms of \textbf{c,e}, with Gaussian broadening. The arrows indicate the directions of the data in \textbf{a,b}, which validate the scenario in \textbf{c,d}.}
\label{Fig1}
\end{figure}
Even more recently, this phenomenology was confirmed in Bi-based materials (with ${Q}^{*}\!\sim\!0.26$ and $0.3$ in single- and double-layer compounds, respectively), following observations in both bulk/momentum space (with RXS) and surface/real space (with STM) \cite{Comin_Science,dSN_Science}, as well as in the electron-doped Nd${}_{2-x}$Ce${}_{x}$CuO${}_{4}$ where ${Q}^{*}\!\sim\!0.24$ was observed by RXS \cite{NCCO_Science}. These multiple experimental observations establish a ubiquitous instability toward charge ordering in the underdoped cuprates.

The microscopic mechanisms that lead to charge order, and govern its interplay with superconductivity and magnetism, are key to the ultimate understanding of the multiple electronic phases that emerge out of the interaction between charge, spin, and lattice degrees of freedom. The relevance of this electronic instability has been extensively pointed out \cite{Kivelson_RMP,Seibold2012} and recently resurged as a prominent topic \cite{Efetov_NatPhys,Sachdev2013,Davis_Lee_2013,Efetov_PRB,Levin_2013,Bulut_2013,Kivelson_2013,DallaTorre_NJP,Lee2014,Chubukov2014,Norman2014},
sparking an intense debate and urging the need for further experimental investigations of the microscopic structure of the charge-ordered state. Several important questions -- such as where charges reside and what is their local symmetry -- were recently addressed at both the theoretical \cite{Efetov_NatPhys,Sachdev2013} and experimental level \cite{Fujita_2014,Achkar_s_wave}.

Here we explore the detailed momentum structure of the charge-density-wave (CDW) order ${\Delta}_{\mathrm{CDW}} (\mathbf{k},\mathbf{Q})$ using RXS, which probes the electronic density directly in reciprocal space, with extreme sensitivity. Our study addresses two major open questions: (i) whether CDW signatures in $({Q}_{x},{Q}_{y})$ space are found exclusively along the reciprocal space directions $({Q}^{*},0)$ and $(0,{Q}^{*})$, or whether they are also present along $({Q}^{*},{Q}^{*})$, as discussed in \cite{Efetov_NatPhys,Sachdev2013,Bulut_2013,Levin_2013,DallaTorre_NJP,Lee2014,Chubukov2014}; (ii) how are charges distributed spatially, and what is the resulting local symmetry of the ordered state \cite{Efetov_NatPhys,Davis_Lee_2013,Sachdev2013,Efetov_PRB,Chubukov2014,Norman2014}. In more general terms, points (i) and (ii) relate to the $\mathbf{Q}$- (\textit{inter-unit-cell}) and $\mathbf{k}$- (\textit{intra-unit-cell}) dependence of the charge order, respectively.
\begin{figure}[t!]
\includegraphics[width=0.85\linewidth]{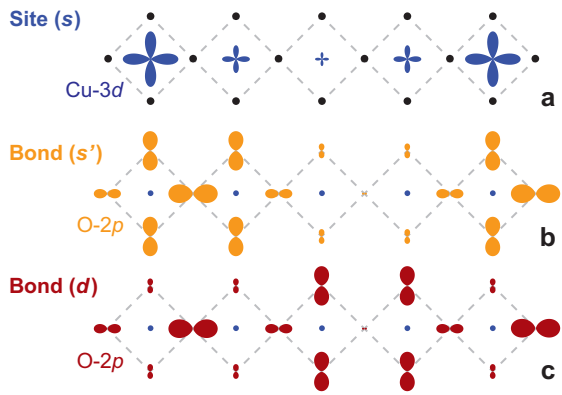}
\caption{{\bf Charge modulation symmetry components}. Real-space schematics of the electronic density $\rho = \bar{\rho} + \delta \rho$ in the case of (\textbf{a}) site-order (charges on Cu), or bond-order (charges on O) with either extended \textit{s}-wave (\textbf{b}) or \textit{d}-wave (\textbf{c}) local symmetry (top to bottom), along a single crystallographic direction.}
\label{Fig2}
\end{figure}
\begin{figure*}[t!]
\includegraphics[width=1\linewidth]{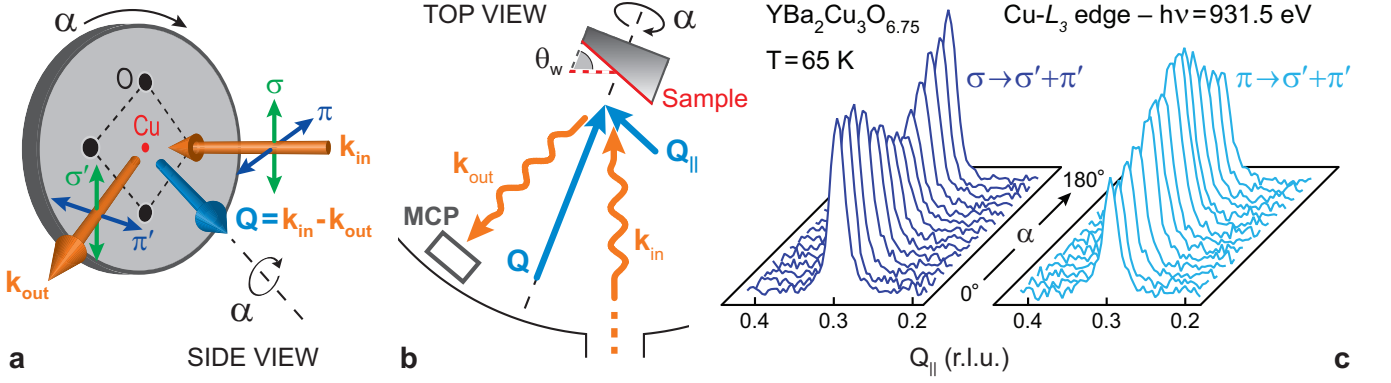}
\caption{{\bf Azimuthal angle-dependent RXS measurements: geometry and experimental data.} \textbf{a}, Side view of the experimental geometry; control variables are: (i) the incoming and outgoing photon wavevectors ${\mathbf{k}}_{\mathrm{in}}$ and ${\mathbf{k}}_{\mathrm{out}}$, which determine the exchanged momentum $\mathbf{Q}$; (ii) the incoming (linear) polarization ${\boldsymbol \epsilon}_{\mathrm{in}}$ ($=\! \sigma$ or $\pi$); (iii) the azimuthal angle $\alpha$, whose rotation axis ${\hat{\mathbf{u}}}_{\alpha}$ coincides with the direction of $\mathbf{Q}$. The polarization of scattered x-rays (${\sigma}^{\prime}$ or ${\pi}^{\prime}$) is not analyzed. \textbf{b}, Top view, illustrating the need for a wedge-shaped sample holder to guarantee the condition $ {\hat{\mathbf{u}}}_{\alpha} \parallel \mathbf{Q}$ for the specific $\mathbf{Q}$-vector of interest (${\theta}_{\mathrm{w}}\!=\!{57.5}^{\circ}$ and ${62}^{\circ}$ for YBCO and Bi2201, respectively). The full (dashed) red line defines the geometry corresponding to $\alpha \!=\! {0}^{\circ}$ ($\alpha \!=\! {180}^{\circ}$). Scattered photons are collected using a multi-channel-plate (MCP) detector. \textbf{c}, Azimuthal angle-dependent $\mathbf{Q}$-scans of the CDW peak (after subtraction of fluorescence background) at ${\mathbf{Q}}_{\mathrm{CDW}}\!=\!(0,0.31,1.5)$ in YBCO-Ortho\,III, plotted vs. the CuO${}_{2}$-plane projection of the exchanged momentum ${Q}_{\parallel}$.}
\label{Fig3}
\end{figure*}

The first part of this work, aimed at addressing the $\mathbf{Q}$-structure of ${\Delta}_{\mathrm{CDW}}$, was performed on the underdoped single-layer compound Bi${}_{2}$Sr${}_{1.2}$La${}_{0.8}$CuO${}_{6+\delta}$ (Bi2201-UD15K), with hole doping $p\!\sim\!0.11$ and ${T}_{\mathrm{c}}\!=\!15$\,K. This \mbox{material} exhibits signatures of incommensurate CDW with wavevectors $({Q}^{*},0)$ and $(0,{Q}^{*})$ (${Q}^{*}\!=\!0.265$) \cite{Comin_Science}. The smaller value of ${Q}^{*}$ allows reaching -- at the Cu-${L}_{3}$ edge -- momenta located near $({Q}^{*},{Q}^{*})$ which in contrast are \textit{not} accessible in double-layer YBCO and Bi${}_{2}$Sr${}_{2}$CaCu${}_{2}$O${}_{8+\delta}$. We use RXS to selectively probe the CuO${}_{2}$-derived electronic states by tuning the photon energy to the Cu-${L}_{3}$ absorption resonance (Fig.\,\ref{Fig1}a). The corresponding experimental results for the momentum-resolved electronic density in the CuO${}_{2}$ planes are shown in Fig.\,\ref{Fig1}b for the two high-symmetry directions $\left( H,0 \right)$ and $\left( H,H \right)$ in the (${Q}_{x}$,${Q}_{y}$) plane. Due to the presence of charge order peaks both along $\left( H,0 \right)$ and $\left( 0,H \right)$, the experimental data are compatible with both checkerboard order (bidirectional) or alternating stripes (unidirectional). In the case of bidirectional order, the two simplest modulation patterns of the charge density $\Delta \rho (x,y)$ with wavevector ${Q}^{*}\!=\!0.265$\,(r.l.u.) are given by: (i) $\Delta \rho (x,y)\!=\!\cos \left( {Q}^{*} x \right) + \cos \left( {Q}^{*} y \right)$ (Fig.\,\ref{Fig1}c); and (ii) $\Delta \rho (x,y)\!=\!\cos \left( {Q}^{*} x \right) \times \cos \left( {Q}^{*} y \right)$ (Fig.\,\ref{Fig1}e). Case (i) corresponds to reciprocal space features along the $\left( H,0 \right)$ and $\left( 0,H \right)$ axes (Fig.\,\ref{Fig1}d), whereas (ii) yields spatial frequencies along the $\left( H,H \right)$ and $\left( H,-H \right)$ direction (Fig.\,\ref{Fig1}f). Similar \textit{Q}-space patterns would be obtained in the case of alternating stripes. Since no CDW peaks are observed along $\left( H,H \right)$, we conclude that the scenario (ii) can be ruled out, thus establishing that charge modulations exclusively run parallel to the Cu-O bond directions (\textbf{a} and \textbf{b} axes). 

The second and main part of this study focuses on the $\mathbf{k}$-structure of the CDW order, which controls the local arrangement of excess charges within each CuO${}_{4}$ plaquette. RXS is able to probe the local charge density $\Delta \rho (\mathbf{r})$ through the spatial modulation of the core (Cu-$2p$) to valence (Cu-$3d$) transition energies $\Delta E (\mathbf{r})$ \cite{achkar2012,achkar2013}. Most importantly, the local symmetry of the valence orbitals (Cu-$3d$ and O-$2p$) is imprinted onto the scattering tensor, which ultimately determines the observed RXS signal (see Supplementary Information for a more detailed derivation). In order to evaluate the symmetry of the CDW order ${\Delta}_{\mathrm{CDW}}$, we selectively probe the different transition channels (Cu-$2{p}_{x,y,z} \rightarrow 3d$) by rotating the light polarization in the RXS measurements. This procedure allows reconstructing the scattering tensor and disentangle the contributions from the different symmetry components of $\Delta_{\mathrm{CDW}} (\mathbf{k},\mathbf{Q}) \!=\! \left\langle {c}^{\dagger}_{\mathbf{k+Q/2}} \cdot {c}_{\mathbf{k-Q/2}} \right \rangle$ \cite{Metlitski2010,Sachdev2013}, namely: (i) a site-centered modulation (${\Delta}_{\mathrm{CDW}}\!=\! {\Delta}_{s}$), corresponding to an extra charge residing on the Cu-$3d$ orbital (Fig.\,\ref{Fig2}a); (ii) an extended \textit{s'}-wave bond-order [${\Delta}_{\mathrm{CDW}}\!=\! {\Delta}_{{s}^{\prime}} (\cos {k}_{x} \!+\! \cos {k}_{y}) $], where the spatially-modulated density is on the O-$2p$ states, and the maxima along the x and y directions coincide (Fig.\,\ref{Fig2}b); (iii) a \textit{d}-wave bond-order [${\Delta}_{\mathrm{CDW}}\!=\! {\Delta}_{d} (\cos {k}_{x} \!-\! \cos {k}_{y}) $], where the charge modulation changes sign between x- and y-coordinated oxygen atoms, and the maxima are shifted by a half wavelength (Fig.\,\ref{Fig2}c).

In the experiments we use a special geometry, in which the sample is rotated around the ordering vector ${\mathbf{Q}}^{*}$ (Fig.\,\ref{Fig3}a,b). This method allows looking at the same wavevector while modulating (as a function of the azimuthal rotation angle $\alpha$) the relative weight of the Cu $2{p}_{x,y,z} \!\rightarrow\! 3{d}$ transitions, which is controlled by the light polarization through dipole selection rules. Here the $\alpha$ dependence of the charge order intensity is the new information that allows evaluating -- through comparison with theoretical predictions from scattering theory -- what is the optimal mix of the \textit{s}-, \textit{s'}-, and \textit{d}-wave symmetry terms that best reproduces the experimental results via their contribution within the scattering tensor. The azimuthal dependence of the RXS signal was studied in Bi2201-UD15K, at ${\mathbf{Q}}^{*} \!\sim\! (0.265,0,2.8)$, and in two underdoped YBa${}_{2}$Cu${}_{3}$O${}_{y}$ compounds: YBa${}_{2}$Cu${}_{3}$O${}_{6.51}$ (YBCO-Ortho\,II, with $p \!\simeq\! 0.10$) and YBa${}_{2}$Cu${}_{3}$O${}_{6.75}$ (YBCO-Ortho\,III, with $p \!\simeq\! 0.13$), at ${\mathbf{Q}}^{*} \!\sim\! (0,0.31,1.5)$.  A series of in-plane momentum (${Q}_{\parallel}$) scans of the charge order peak in YBCO-Ortho\,III, acquired at $T \!=\! {T}_{\mathrm{c}}\!=\!75$\,K with both $\sigma$- and $\pi$-polarized incoming X-rays, is presented in Fig.\,\ref{Fig3}c for the range ${0}^{\circ}<\alpha<{180}^{\circ}$, where $ \alpha \!=\! 0$ corresponds to having the \textbf{b} axis in the scattering plane (as determined by high-energy Bragg diffraction) in the configuration of Fig.\,\ref{Fig4}b (full red line). 

The total scattered intensity ${I}_{\mathrm{RXS}}$ is extracted by fitting the RXS momentum scans with a Gaussian peak, and is in general proportional to the amplitude of the charge modulation. We can directly compare ${I}_{\mathrm{RXS}}$ to the theoretical scattering tensor ${F}_{pq}$ \cite{Tjeng_2005,DiMatteo_2012}:
\begin{figure*}[t!]
\includegraphics[width=1\linewidth]{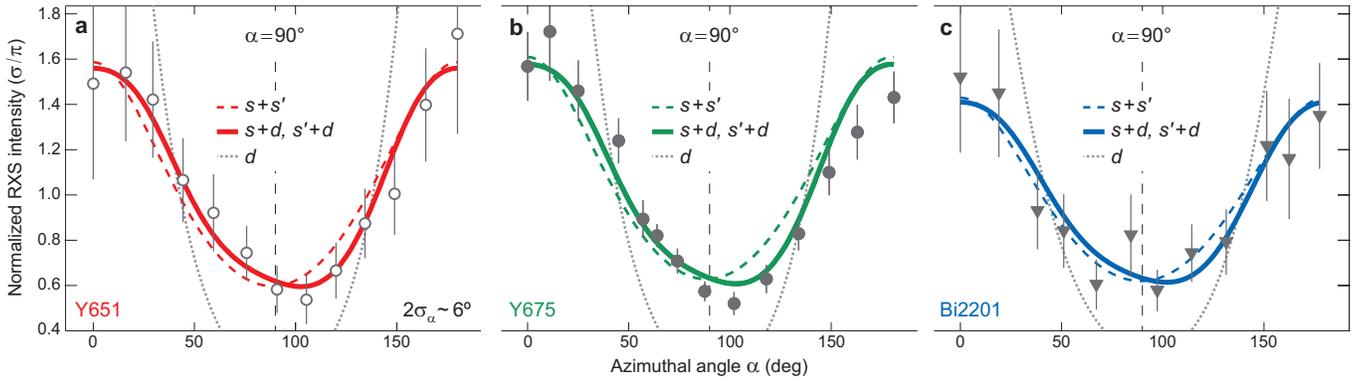}
\caption{{\bf Experimental and calculated CDW peak intensity vs. azimuthal angle.} Normalized RXS intensity ratio $ {I}_{\mathrm{RXS}}^{\sigma} / {I}_{\mathrm{RXS}}^{\pi} $ for Y651 (\textbf{a}); Y675 (\textbf{b}); and Bi2201 (\textbf{c}).  The vertical error bars are obtained by propagation of the uncertainties for the $\sigma$ and $\pi$ RXS intensities as obtained from a non-linear least-squares regression analysis of the raw experimental data. Theoretical profiles for three possible two-component combinations are obtained from a least-squares fitting method and overlaid to the data: $s + d$ and ${s}^{\prime} + d$ (full, a single trace is used since the resulting profiles are nearly overlapping), and $s + {s}^{\prime}$ (dashed). Best-fit, pure $d$-wave azimuthal profiles are also shown as dashed grey curves. For all data points the uncertainty on the azimuthal angle is given as ${\sigma}_{\alpha} \sim {3}^{\circ}$. The vertical dashed lines mark the azimuthal position $\alpha \!=\! {90}^{\circ}$.}
\label{Fig4}
\end{figure*}
\begin{equation}
{I}_{\textstyle {\boldsymbol \epsilon} \!\rightarrow\! {\boldsymbol \epsilon}^{\prime}} \left( \mathbf{{Q}^{*}} , \alpha \right) \propto {\left\vert {\sum}_{p q} {\epsilon}_{p} \cdot {F}_{pq} \left( \mathbf{{Q}^{*}}, \alpha \right) \cdot {\epsilon}^{\prime}_{q} \right\vert}^{2}
\end{equation}
where $\boldsymbol \epsilon$ and ${\boldsymbol \epsilon}^{\prime}$ represent the polarization vectors for incoming and outgoing photons, respectively, while $ \mathbf{{Q}^{*}} $ is the ordering wavevector. The $\alpha$ dependence is induced by simply applying a rotation (about the azimuthal axis and of magnitude equal to $\alpha$) to the scattering tensor ${F}_{pq} $. Based on symmetry arguments (see Supplementary Information for additional details) the unrotated scattering tensor ${F}_{pq} $ can be written in terms of a linear combination of the $s$-, ${s}^{\prime}$-, and $d$-wave components of the charge order, with respective magnitudes ${\delta}_{s}$, ${\delta}_{s'}$, and ${\delta}_{d}$ [note that, since the scattering yield at the Cu-${L}_{3}$ edge is more sensitive to charges on the Cu site ($s$-wave order) than on the O site (${s}^{\prime}$- and $d$-wave order), we have that ${\delta}_{d} / {\delta}_{s'} \!=\! {\Delta}_{d} / {\Delta}_{s'}$, while in general ${\delta}_{s} / {\delta}_{d} > {\Delta}_{s} / {\Delta}_{d}$ and ${\delta}_{s} / {\delta}_{s'} > {\Delta}_{s} / {\Delta}_{s'}$; see Supplementary Information for more details]. This way ${F}_{pq} $ becomes:
\begin{equation}
{F}_{pq} \left( \pm \mathbf{{Q}^{*}} \right) = \left\vert \begin{array}{ccc} {\delta}_{s} + \left( {\delta}_{s'} + {\delta}_{d} \right) \cos \phi & 0 & 0 \\ 0 & {\delta}_{s} + {\delta}_{s'} - {\delta}_{d} & 0 \\ 0 & 0 & \gamma {\delta}_{s} \end{array} \right\vert
\label{Scattering_tensor_combination}
\end{equation}
where the phase $\phi \!=\! {Q}^{*} \cdot a/2 $ accounts for the mismatch between the ordering period and the lattice parameter, while $\gamma$ is the ratio between the out-of-plane and the in-plane transition matrix elements, which has been estimated from x-ray absorption data on Bi2201 (a similar analysis in YBCO is hampered by the proximity between the chain and plane transitions in the absorption spectrum). Note that a similar version of Eq. \ref{Scattering_tensor_combination}, developed here for the Cu-\textit{L} edge, has been recently used in Ref. \onlinecite{Achkar_s_wave} for RXS at the O-\textit{K} edge.

The total calculated scattering intensity, before self-absorption correction, is then given by: $ {I}_{\mathrm{calc}} \left( \alpha \right) = {I}_{\epsilon \rightarrow {\sigma}^{\prime}} \left( \alpha \right) + {I}_{ \epsilon \rightarrow {\pi}^{\prime}} \left( \alpha \right) $, where $ \epsilon\!=\! \sigma $ or $\pi$. We subsequently include self-absorption corrections on the calculated profiles (see Supplementary Information). Figure\,\ref{Fig4} presents the experimental data for the two YBCO samples and for Bi2201 in the form of the RXS intensity ratio between vertical and horizontal polarization configurations $ {I}_{\mathrm{RXS}}^{\sigma} / {I}_{\mathrm{RXS}}^{\pi} $ (grey markers) in order to factor out possible extrinsic effects due to the sample shape and orientation with respect to the scattering geometry. Also shown are model calculations (${I}_{\mathrm{calc}}$, continuous lines) for all possible combinations of two CDW symmetry components, i.e. $s + d$, ${s}^{\prime} + d$, and $s + {s}^{\prime}$, together with the pure $d$-wave model for comparison (for a complete analysis of all possible combinations of one- and two-symmetry terms see Supplementary Materials). In particular, the peculiarity of those combinations including a $d$-wave term is that the minimum in the calculated profile $ {I}_{\mathrm{calc}}^{\sigma} / {I}_{\mathrm{calc}}^{\pi} $ is displaced from $\alpha \!=\! 90^{\circ}$, a consequence of the more strongly asymmetric pattern of charges within each CuO${}_{4}$ plaquette (see again Fig.\,\ref{Fig2}). On the contrary, a combination of \textit{s} and \textit{s'} components alone remains symmetric with respect to $\alpha \!=\! {90}^{\circ}$, and so do the pure-symmetry profiles. Since the experimental data are characterized by a slight asymmetry (${\alpha}_{\mathrm{min}} \!\simeq\! {100}^{\circ} $), the two-component combinations involving a locally asymmetric ($d$-wave) term fit the YBCO data more closely. For such combinations, the presence of a symmetric term is also found to be necessary, as a pure $d$-wave fit clearly overestimates the total amplitude of the experimental azimuthal modulation (see dashed grey line in Fig.\,\ref{Fig4}). 

On the other hand, the lack of a clear asymmetry in Bi2201 prevents our analysis from providing a conclusive answer on the symmetry of charge order in this material. However, such an asymmetry might be overshadowed by the larger scatter in the data due to weaker CDW features in RXS data on Bi-cuprates than in YBCO. Indeed, we note that this has been assessed -- for (bilayer) Bi-based cuprates -- using alternative approaches \cite{Fujita_2014}.
\begin{table}[b!]
\begin{center}
\begin{tabular}{ccccc}
\hline
\hline
\multirow{3}{*}{\qquad Sample \qquad} & Order & $ s + {s}^{\prime} $ & $ s + d $ & $ {s}^{\prime} + d $ \\
& & & \\
& Ratio & $ {s}^{\prime} / s $ & $ s /d $ & $ {s}^{\prime} / d $ \\
\hline
Y651 & & 0.01 & 0.21 & 0.27 \\
Y675 & & -0.01 & 0.22 & 0.27 \\
\hline
Probability level $P$ & & 5.6 & \textbf{83.8} & \textbf{85.5} \\
\hline
\hline
\end{tabular}
\label{table_report}
\end{center}
\caption{{\bf Statistical comparison of CDW models}. Best-fit component ratios $ {s}^{\prime} / s $, $ s / d $, and $ {s}^{\prime} / d $ for binary combinations of the fundamental CDW symmetry terms $s + {s}^{\prime}$, $s + d$, and ${s}^{\prime} + d$, respectively. Probability levels $P$ for the hypothesis that each specific CDW model fits the experimental data better than a random sample. The values suggest that those combinations featuring a prominent \textit{d}-wave bond-order component manifest a great likelihood ($P \!>\! 90\%$) of reproducing the experimental data. }
\label{Cumulative_probability}
\end{table}

The qualitative argument based on the data asymmetry is supported by a more quantitative assessment of the likelihood of each model, which was estimated by evaluating the reduced chi-square (${\chi}^{2}_{\mathrm{red}}$) for all the experimental points and theoretical configurations shown in Fig.\,\ref{Fig4} (see Supplementary Information for a formal definition of ${\chi}^{2}_{\mathrm{red}}$). The values of ${\chi}^{2}_{\mathrm{red}}$ are subsequently used within the chi-squared cumulative distribution function to extract the probability $P$ for the different models considered here, where \textit{P} denotes the probability that the model under consideration yields a better agreement than a dataset randomly generated from a normal distribution (with mean-square deviations equal to the experimental uncertainties). These probability levels (Table\,\ref{Cumulative_probability}) indicate that a symmetry decomposition including a dominant $d$-wave bond-order component is more likely to describe the experimental data from YBCO than a combination of symmetric $s$- and ${s}^{\prime}$-wave components. Although the relative magnitude of the $d$- versus $s$- or ${s}^{\prime}$-wave character is here not strongly constrained, we note the presence of a symmetric component of about 20\% of the total charge order (see Table\,\ref{table_report} and Supplementary Materials for a more detailed discussion on the analysis); this closely follows theoretical predictions for ${\Delta}_{\mathrm{CDW}}$ in the context of the \textit{t}-\textit{J} model \cite{Vojta_2008,Metlitski2010,Sachdev2013}, as well as recent STM results \cite{Fujita_2014}. Finally, we also note the close proximity between a mixed solution with prevailing d-wave character and those with prevailing $s$- or ${s}^{\prime}$-wave character; this is illustrated in Fig.\,S6, which however indicates that even in the latter case the $d$-wave component would still be as large as 20-30\%. 

Altogether, in YBCO we reveal the charge-ordered electronic ground state to be best described by a bond-order with the modulating charge mainly located on O-2\textit{p} orbitals and characterized by a prominent \textit{d}-wave character, while in Bi2201 the absence of charge order features along the diagonal axes in momentum space demonstrates that charge modulations propagate exclusively along the \textbf{a} and \textbf{b} axes. Therefore, our study reaffirms the pivotal role played by the O-2\textit{p} ligand states in hole-doped cuprates \cite{ZSA,Emery_Reiter}. In light of STM works pointing to bond-order in Ca${}_{1.88}$Na${}_{0.12}$CuO${}_{2}$Cl${}_{2}$, and Bi${}_{2}$Sr${}_{2}$CaCu${}_{2}$O${}_{8+\delta}$ \cite{koshaka2007,lawler2010}, and more recently revealing a dominant $d$-wave symmetry \cite{Fujita_2014}, we propose that in the Bi-, Y-, and Cl-based cuprates, which all exhibit a very similar charge order phenomenology, the microscopic defining symmetry contains a prominent $d$-wave bond-order component. In the La-based cuprates, which already display a doping dependence for the charge ordering vectors opposite to the one of Bi2201 and YBCO \citep{Blackburn2013}, a recent detailed study has revealed a predominant \textit{s'}-wave bond-order \cite{Achkar_s_wave}, suggesting a different manifestation of the charge order symmetry in these systems. In such context, we anticipate that future work will be needed to provide further experimental constraints to the ratio of different symmetry terms, to understand the sensitivity of different probes to the symmetry of the charge order, and possibly also how the latter is modulated by the out-of-plane component of the wavevector.

The commonality between the symmetry of the superconducting (SC) and CDW orders might suggest that the same attractive interaction responsible for particle-particle (Cooper) pairing might also be active in the particle-hole channel. This aspect -- which has been recently proposed at the theoretical level and was suggested to originate from the exchange part (\textit{J}) of the interaction Hamiltonian \cite{Vojta_2008,Metlitski2010,Davis_Lee_2013,Sachdev2013} -- is here corroborated by our experiments. This has deep implications in the context of the competing instabilies of the electronic system and for the ultimate understanding of the pairing mechanism.\\

\noindent {\bf Methods}

\vspace{1mm}\footnotesize{\noindent {\bf Sample characterization.} This study focuses on two underdoped YBa$_{2}$Cu$_3$O$_{6+y}$ single crystals ($y\!=\!0.51$, $p\!\simeq\!0.10$, ${T}_{\mathrm{c}}\!=\!57$\,K, YBCO-Ortho\,II; $y\!=\!0.75$, $p\!\simeq\!0.13$, ${T}_{\mathrm{c}}\!=\!75$\,K, YBCO-Ortho\,III) and one underdoped crystal of Bi${}_{2}$Sr${}_{1.2}$La${}_{0.8}$CuO${}_{6+\delta}$ ($p \!\sim\! 0.11$, ${T}_{\mathrm{c}}\!=\!15$\,K, Bi2201-UD15K). The superconducting critical temperature $T_{\mathrm{c}}$ was determined from magnetic susceptibility measurements. The $T_{\mathrm{c}}$-to-doping correspondence is taken from Ref.\,\onlinecite{Liang_2006} (YBCO) and Ref.\,\onlinecite{Ando2000} (Bi2201).

\noindent {\bf Soft X-ray scattering.} The scattering measurements were performed at beamline REIXS of the Canadian Light Source, on a 4-circle diffractometer in a ${10}^{-10}$\,mbar ultra-high-vacuum chamber, with a photon flux around $5 \cdot {10}^{12}$\,photons/s and $\frac{\Delta E}{E}\!\sim\! 2 \cdot {10}^{-4}$ energy resolution. In addition, fully polarized incoming light is used, with two available configurations: $\sigma$ (polarization vector perpendicular to the scattering plane) or $\pi$ (polarization vector in the scattering plane). Due to poor performance of polarization analyzers in the soft X-ray regime, the polarization of the scattered light was not resolved in any of the measurements. In order to maximize the charge order signal, all measurements were taken at the peak energy of the Cu-${L}_{3}$ edge ($h \nu \!=\! 931.5$\,eV), and at the superconducting transition temperature ${T}_{\mathrm{c}}$. The azimuthal angle $\alpha$ is defined as the angle between the RXS scan direction in the $\left( {Q}_{x}, {Q}_{y} \right)$ plane of momentum space, and the crystallographic \textbf{b} axis (for more details on the azimuthal sample geometry see Fig.\,S1 and corresponding discussion in the Supplementary Information). Note that at all azimuthal angles, the sample tilt angle has been slightly readjusted to ensure that the RXS scans slice across the maximum of the CDW peak.

\providecommand{\noopsort}[1]{}\providecommand{\singleletter}[1]{#1}%

\subsection{Acknowledgments}

\normalsize{We are grateful to J.E. Hoffman, Yang He and M. Yee for sharing their STM data and for fruitful discussions. We also acknowledge M. Le Tacon, S. Sachdev, M. Norman, S. Kivelson, C. Pepin, E. Dalla Torre, and E. Demler for insightful discussions. This work was supported by the Max Planck -- UBC Centre for Quantum Materials, the Killam, Alfred P. Sloan, Alexander von Humboldt, and NSERC's Steacie Memorial Fellowships (A.D.), the Canada Research Chairs Program (A.D., G.A.S.), NSERC, CFI, and CIFAR Quantum Materials. Part of the research described in this paper was performed at the Canadian Light Source, which is funded by the CFI, NSERC, NRC, CIHR, the Government of Saskatchewan, WD Canada, and the University of Saskatchewan. R.C. acknowledges the receipt of support from the CLS Graduate Student Travel Support Program. E.H.d.S.N. acknowledges support from the CIFAR Global Academy.}

\subsection{Author Contributions}

R.C., B.K., G.A.S., and A.D. conceived this investigation -- R.C. performed RXS measurements at Canadian Light Source with the assistance of R.S., F.H., E.H. d.S.N., and L.C. -- R.C. developed the theoretical model and performed related calculations -- R.C., A.F., A.J.A., D.G.H., B.K., G.A.S., and A.D. are responsible for data analysis and interpretation -- R.L., W.N.H. and D.B. provided the YBCO samples -- Y.Y and H.E. provided the Bi2201 samples. All of the authors discussed the underlying physics and contributed to the manuscript. R.C. and A.D. wrote the manuscript. A.D. is responsible for overall project direction, planning, and management.

\clearpage

\onecolumngrid

\makeatletter
\renewcommand{\fnum@figure}{\figurename~S\thefigure}
\setcounter{figure}{0}
\makeatother
\renewcommand{\thetable}{S\arabic{table}}
\renewcommand{\theequation}{S\arabic{equation}} 

\vspace{6mm}

\begin{center}
\Huge{Supplementary Information}

\vspace{6mm}

\Large{\textbf{The symmetry of charge order in cuprates}}

\vspace{4mm}

\large{R. Comin,$^{\ast,1,2}$ R. Sutarto,$^{3}$ F. He,$^{3}$ E. H. da Silva Neto,$^{1,2,4,5}$ L. Chauviere,$^{1,2,4}$\\A. Frano,$^{5,6}$ R. Liang,$^{1,2}$ W. N. Hardy,$^{1,2}$ D. Bonn,$^{1,2}$ Y. Yoshida,$^{7}$ H. Eisaki,$^{7}$\\A.J.\,Achkar,$^{8}$ D.G.\,Hawthorn,$^{8}$ B. Keimer,$^{4}$ G. A. Sawatzky,$^{1,2}$\\and A. Damascelli$^{\ast,1,2}$}

\vspace{4mm}

\title{The nature of charge ordering in cuprates}

\vspace{6mm}

\normalsize{$^{1}$Department of Physics {\rm {\&}} Astronomy, University of British Columbia, Vancouver, British Columbia V6T\,1Z1, Canada}\\
\vspace{2mm}\normalsize{$^{2}$Quantum Matter Institute, University of British Columbia, Vancouver, British Columbia V6T\,1Z4, Canada}\\
\vspace{2mm}\normalsize{$^{3}$Canadian Light Source, Saskatoon, Saskatchewan  S7N 2V3, Canada}\\
\vspace{2mm}\normalsize{$^{4}$Max Planck Institute for Solid State Research, Heisenbergstrasse 1, D-70569 Stuttgart, Germany}\\
\vspace{2mm}\normalsize{$^{5}$Quantum Materials Program, Canadian Institute for Advanced Research, Toronto, Ontario M5G 1Z8, Canada}\\
\vspace{2mm}\normalsize{$^{6}$Helmholtz-Zentrum Berlin f\"{u}r Materialien und Energie, Wilhelm-Conrad-R\"{o}ntgen-Campus BESSY II, Berlin, Germany}\\
\vspace{2mm}\normalsize{$^{7}$National Institute of Advanced Industrial Science and Technology (AIST), Tsukuba, 305-8568, Japan}\\
\vspace{2mm}\normalsize{$^{8}$Department of Physics and Astronomy, University of Waterloo, Waterloo, N2L 3G1, Canada}\\

\vspace{4mm}

\normalsize{$^\ast$To whom correspondence should be addressed.}\\
\normalsize{E-mail: r.comin@utoronto.ca (R.C.), damascelli@physics.ubc.ca (A.D.).}

\end{center}







\newpage

\noindent {\bf Azimuthal sample geometry.}
\\
Our study relies on the capability of rotating the sample crystallographic axes with respect to a given ordering wavevector $\mathbf{Q}$, and subsequently slicing across the ordering peak in momentum space along different directions in the $\left( {Q}_{x}, {Q}_{y} \right)$ plane.

In order to implement this experimental scheme, we need to establish a geometry for the sample holder which allows to rotate the sample around an axis coinciding with the transferred momentum (which, in our case, also coincides with the ordering wavevector). However, if we mount the sample on a conventional (flat) sampleholder (SH), i.e. with the crystallographic \textbf{a-b}-plane coincident with the basal plane of the SH, the transferred momentum will be parallel to the out-of-plane wavevector ${Q}_{z}$, with zero projection to the $\left( {Q}_{x}, {Q}_{y} \right)$ plane. Therefore, in order to reach the charge order reflection at ${Q}_{y} \!\sim\! 0.31$ reciprocal lattice units we use a wedge-shaped SH (see Fig.\,3b in the main text), which allows offsetting the sample crystallographic \textbf{b}-axis to an amount functional to reach the desired position in the $\left( {Q}_{x}, {Q}_{y} \right)$ plane. 
Such configuration, with the \textbf{b}-axis rotated but still in the scattering plane, corresponds to the azimuthal angle $\alpha \!=\! {0}^{\circ}$ in our definition. This situation is illustrated in Fig.\,S\ref{Azimuthal_geometry}a1, which clarifies how the offset in the sample orientation induces a nonzero planar projection (${\mathbf{Q}}_{\parallel}$) of the wavevector $\mathbf{Q}$. The top and side views of this configurations are shown in Fig.\,S\ref{Azimuthal_geometry}a2, while the schematic in Fig.\,S\ref{Azimuthal_geometry}a3 shows the location of the wavevector ${\mathbf{Q}}_{\parallel}$ in the $\left( {Q}_{x}, {Q}_{y} \right)$ plane, as well as the direction of the momentum scan (see green box) when the sample is rotated in the scattering plane. This scheme elucidates how the $\alpha \!=\! {0}^{\circ}$ azimuthal geometry corresponds to performing the momentum scan across the ordering peak at ${\mathbf{Q}}_{b} \sim \left( 0, 0.31, L \right)$ (in the case of YBCO) along a direction parallel to the ${Q}_{y}$ axis, where $L$ is the out-of-plane component of the ordering wavevector (for this study, we used $L \simeq 1.5$).
By changing the azimuthal angle, the sample revolves around the axis parallel to the transferred momentum $\mathbf{Q}$, and the corresponding configurations for $\alpha \!=\! {90}^{\circ}$, ${180}^{\circ}$, and ${270}^{\circ}$ are shown in Figs.\,S\ref{Azimuthal_geometry}b1-b3, Figs.\,S\ref{Azimuthal_geometry}c1-c3, and Figs.\,S\ref{Azimuthal_geometry}d1-d3, respectively. In particular, from the diagrams in Figs.\,S\ref{Azimuthal_geometry}a3, b3, c3, and d3, one can note how the projection of the central value in the momentum scan always remains the same (dark red arrows), a consequence of the fact that the azimuthal rotation leaves the ordering wavevector invariant since the latter coincides with the azimuthal axis of rotation. However, the direction of the momentum scan is now rotated with respect to the ${Q}_{x}$ and ${Q}_{y}$ axes, thus realizing the requirement necessary to perform this study.
\begin{figure}[t!]
\centering
\includegraphics[width=1\linewidth]{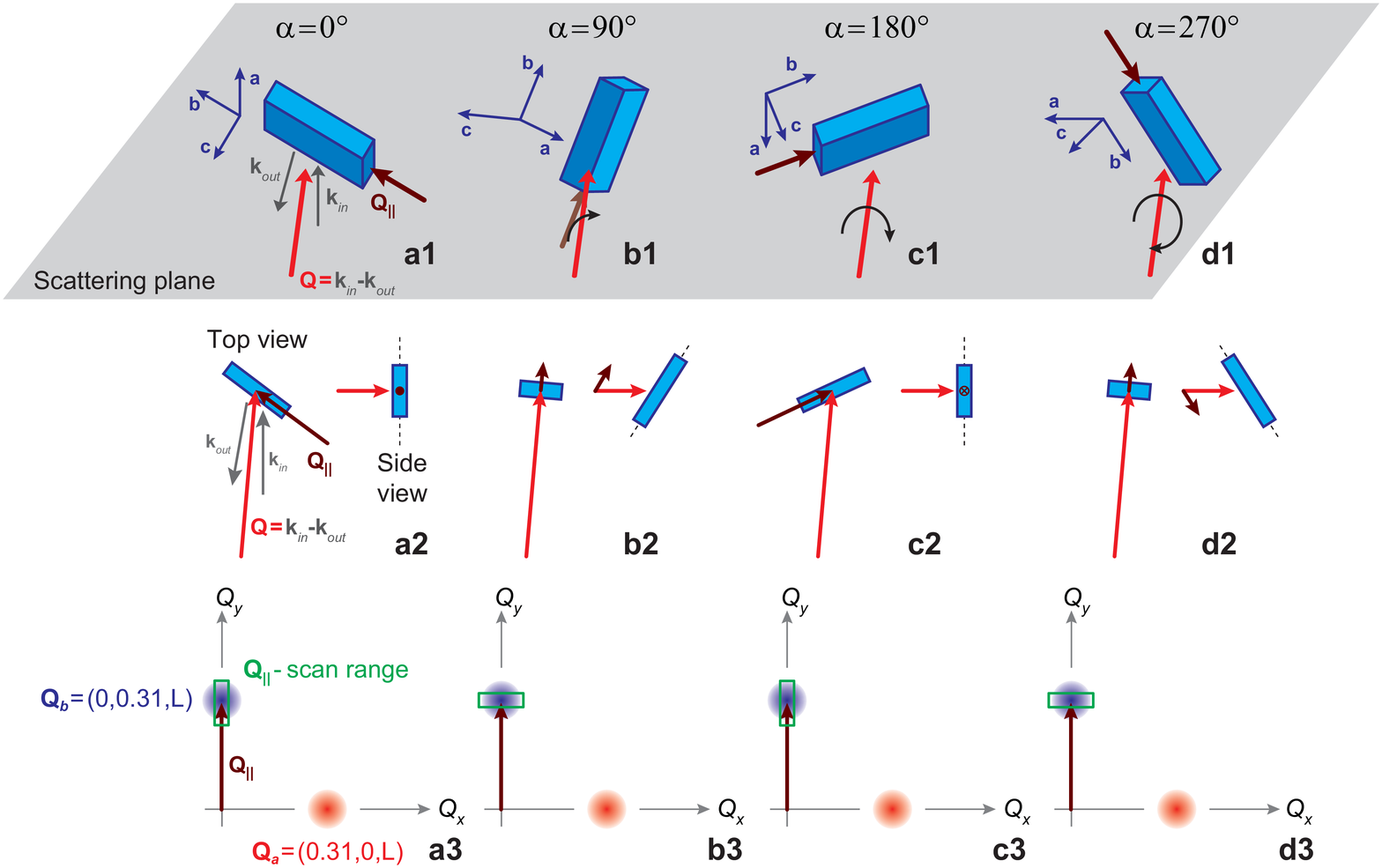}
\caption{{\bf Schematics of sample geometry implementing the azimuthal rotation.} \textbf{a1,b1,c1,d1}, Projected views of the orientation of the sample crystallographic frame with respect to the scattering plane and the transferred momentum $\mathbf{Q}$ for the case of azimuthal angles $\alpha \!=\! {0}^{\circ}$, ${90}^{\circ}$, ${180}^{\circ}$, and ${270}^{\circ}$, respectively. \textbf{a2,b2,c2,d2}, Top and side views of the configurations in \textbf{a1,b1,c1,d1}. \textbf{a3,b3,c3,d3}, Projections of the transferred momenta and scan directions in the ${Q}_{x}$, ${Q}_{y}$ scattering plane, for the corresponding azimuthal angle values.}
\label{Azimuthal_geometry}
\end{figure}
\\

\noindent {\bf Polarization-dependent X-ray absorption.}
\\
The photon energy ($\omega$) and site ($n$) dependent form factor ${f}_{p q}^{(n)} \!\left( \omega \right)$ encodes all the information that can be experimentally retrieved using X-ray absorption (XAS) and scattering (RXS), and is mathematically defined as follows:
\begin{equation}
{f}_{p q}^{(n)} \!\left( \omega \right) = \frac{{e}^{2}}{\hbar {m}^{2} {c}^{2}} {\vert \mathbf{A} \vert}^{2}  \: {\sum}_{i,f} \, \frac{ \langle {\psi}_{i}^{(n)} \vert {\mathrm{p}}_{q} \vert {\psi}_{f}^{(n)} \rangle \cdot \langle {\psi}_{f}^{(n)} \vert {\mathrm{p}}_{p} \vert {\psi}_{i}^{(n)} \rangle}{\omega - ({\omega}_{f}^{(n)}-{\omega}_{i}^{(n)})+i{\Gamma}_{if}}
\end{equation}
\noindent
where $e$ and $m$ are the fundamental electronic charge and mass, $\mathbf{p}\!=\!{\left\lbrace {\mathrm{p}}_{p} \right\rbrace}_{p=x,y,z}$ is the electron momentum operator, and $\mathbf{A}$ is the electromagnetic vector potential. Here ${\psi}_{i}^{(n)}$ and ${\psi}_{f}^{(n)}$ represent the initial and final single-particle electronic states at site ${\mathbf{R}}_{n}$ (with energies ${\omega}_{i}^{(n)}$ and ${\omega}_{f}^{(n)}$, respectively) involved in the light-induced transition $i \rightarrow f$. ${\Gamma}_{if}$ is the lifetime of the intermediate state with an electron in ${\psi}_{i}^{(n)}$ and a hole in ${\psi}_{f}^{(n)}$. Henceforth we will use the unit vectors ${\boldsymbol \epsilon}$ and ${{\boldsymbol \epsilon}}^{\prime}$ to refer to the polarization state (direction of the vector potential $\mathbf{A}$) of incoming and outgoing photons.

\noindent
The observables associated to XAS and RXS techniques are directly related to ${f}_{pq}^{(n)} \left( \omega \right)$ [1]:
\begin{eqnarray}
{I}^{\mathrm{XAS}} \left( \omega \right) \!& \propto &\! - \frac{1}{{\omega}^{2}} \times \mathrm{Im} \left[ {\sum}_{n} {\sum}_{p} \: {\epsilon}_{p} \cdot {f}_{pp}^{(n)} \left( \omega \right) \right]
\label{XAS_def}
\\
{I}^{\mathrm{RXS}} \left( \mathbf{Q}, \omega \right) \!& \propto &\! {\left\vert {\sum}_{p q} {\epsilon}_{p} \cdot \left[ {\sum}_{n} \: {f}_{pq}^{(n)} \left( \omega \right) {e}^{i \mathbf{Q} {\textstyle\cdot} \: {\mathbf{R}}_{n}} \right] \cdot {\epsilon}^{\prime}_{q} \right\vert}^{2} = {\left\vert {\sum}_{p q} {\epsilon}_{p} \cdot {F}_{pq} \cdot {\epsilon}^{\prime}_{q} \right\vert}^{2}
\label{RXS_def}
\end{eqnarray}
where we have introduced the scattering tensor ${F}_{pq}$, which \textit{is not} a local quantity (does not depend on the lattice position ${\mathbf{R}}_{n}$) and is more directly related to the physical observable in RXS experiments (${I}^{\mathrm{RXS}}$). We note that an equivalent approach was shown in [2], but there $F_{pq}$ is denoted $T$. Moreover, from the above equations it follows that XAS only depends on the incoming light polarization ${\epsilon}_{p}$, whereas the RXS signal depends on the outgoing light polarization ${\epsilon}^{\prime}_{q}$, as well.

first of all, the local form factor inherits the symmetry properties of the material-specific space group. For a non-magnetic orthorhombic system, and assuming the Cartesian axes $\mathbf{x},\mathbf{y},\mathbf{z}$ to coincide with the crystallographic axes $\mathbf{a},\mathbf{b},\mathbf{c}$, one has:
\begin{equation}
{f}_{pq} = \left\vert \begin{array}{ccc}
{f}_{xx} & 0 & 0 \\
0 & {f}_{yy} & 0 \\
0 & 0 & {f}_{zz} \end{array} \right\vert,
\end{equation}
with ${f}_{xx}\!\neq\!{f}_{yy}\!\neq\!{f}_{zz}$, in general.

\noindent
YBCO and Bi2201 are both orthorhombic materials, but the origin of their orthorhombicity is different. In Bi2201, the orthorhombic distortion consists of a tiny rhomboedral deformation of the structural unit cell along the ${\mathbf{b}}^{*}$ axis, oriented at 45 degrees from the Cu-O bond direction [3,4,5]. Although the CuO${}_{2}$ planes cease to have square symmetry, the effective anisotropy between the two planar axes $\mathbf{a}$ and $\mathbf{b}$ is so tiny that one can approximate ${f}_{xx}\!\simeq\!{f}_{yy}\!\neq\!{f}_{zz}$. On the other hand, in YBCO the orthorhombicity originates from the presence of the chain layer, where the partially-oxygenated Cu-O chains run along the $\mathbf{b}$ axis, thus making $\mathbf{a}$ and $\mathbf{b}$ inequivalent even though the CuO${}_{2}$ planes formally retain square symmetry on their own. In principle, near the Cu absorption edges, the form factor can decomposed as ${f}_{pq}^{\mathrm{Cu}}\!=\!{f}_{pq}^{\mathrm{plane}} + {f}_{pq}^{\mathrm{chain}}$, with ${f}_{xx}^{\mathrm{plane}}\!=\!{f}_{yy}^{\mathrm{plane}}$ and ${f}_{xx}^{\mathrm{chain}}\!\neq\!{f}_{yy}^{\mathrm{chain}}$. Unfortunately, the plane- and chain-related features overlap at the Cu-${L}_{2,3}$ edge in YBCO, and therefore cannot be fully disentangled [6]. For these reasons, we have elected to study the polarization-dependence in the XAS on Bi2201, whose doping lies very close to the YBCO samples, in order to extract a reliable estimate for the diagonal elements in the scattering tensor. The latter constitute a crucial experimental input for the model later employed to calculate the azimuthal angle dependent RXS cross-section.

The photon energy dependence of ${f}_{pq}^{(n)} \left( \omega \right)$ near the Cu-${L}_{2,3}$ absorption can be modeled by a simple Lorentzian lineshape, since no multiplet structure is present. In general, the parameters of this Lorentzian function (amplitude $A$; position $\Delta E$; linewidth $\Gamma$) are site-dependent, so that we can write, in the most general case:
\begin{equation}
{f}_{p p}^{(n)} (\omega) \sim {A}^{(n)} {\left( \omega - \Delta {E}_{p}^{(n)} + i{\Gamma}^{(n)} \right)}^{-1}
\label{Scattering_tensor_single}
\end{equation}
\noindent
$\Gamma$ is inversely proportional to the lifetime of the $3d - 2p$ electron-hole excitation, therefore it is hardly affected by small spatial variations of the electronic density, and we can set ${\Gamma}^{(n)}\!=\!\Gamma$. On the other hand, the amplitude $A$ and peak position $\Delta E$, depending on the local density of unoccupied states and on the energy of initial/final state respectively, might vary as a function of lattice position as a consequence of the modulated charge density. However, in the cuprates, it has been shown that spatial variations of the transition amplitude $A$ are not the main mechanism behind the photon energy-dependent RXS response [7,8], hence we also assume ${A}^{(n)} \rightarrow A$. The site-dependent transition energies depend on the charge density, and will therefore also reflect the symmetry of the charge-ordered state. They can be readily calculated from the charge-density-wave (CDW) order ${\Delta}_{\mathrm{CDW}}$, which is discussed in the next section.
\begin{figure}[b!]
\centering
\includegraphics[width=0.7\linewidth]{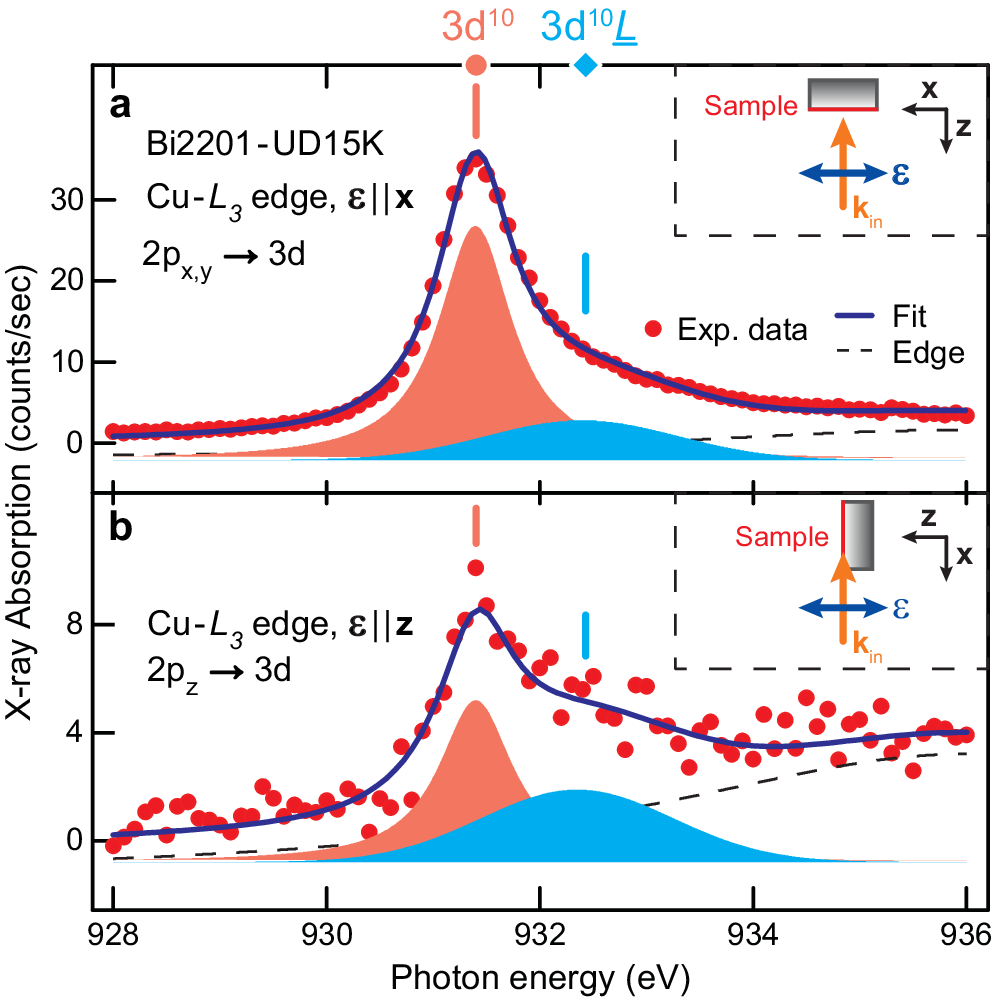}
\caption{{\bf Polarization-dependent XAS in Bi2201}. \textbf{a}, X-ray absorption profile at the Cu-${L}_{3}$ edge using in-plane light polarization (${\boldsymbol \epsilon} \!\parallel\! \mathbf{x}$), which allows accessing Cu-$2{p}_{x,y} \rightarrow 3{d}_{{x}^{2}-{y}^{2}}$ transitions for sites with a $3{d}^{9}$ or $3{d}^{9} \underline{L}$ configuration. Inset: sketch of the experimental geometry for ${\boldsymbol \epsilon}\!\parallel\! \mathbf{x}$. \textbf{b}, same as \textbf{a}, but with out-of-plane light polarization (${\boldsymbol \epsilon} \!\parallel\! \mathbf{z}$), where Cu-$2{p}_{z} \rightarrow 3{d}_{3{z}^{2}-{r}^{2}}$ transitions can be measured.}
\label{XAS}
\end{figure}
For what concerns the polarization dependence at the Cu-${L}_{3}$ edge, the cross section for different incoming light polarization varies according to the orbital character of the initial and final states. The three possible configurations ${\boldsymbol \epsilon} \!\parallel\! \mathbf{x},\mathbf{y},\mathbf{z}$ only allow (in the dipole approximation) excitation of a core electron from a Cu-$2{p}_{x}$, $2{p}_{y}$ or $2{p}_{z}$ orbital, respectively. For a hole-doped CuO${}_{2}$ plane, two final states can be reached in the excitation process: (i) a Cu-$2{p}^{5} 3{d}^{10}$ state, with filled Cu-$3d$ and O-$2p$ shells; or (ii) a Cu-$2{p}^{5} 3{d}^{10}\underline{L}$ configuration, where a ligand hole is present. Both final states have a nonzero Cu-$3{d}_{{x}^{2}-{y}^{2}}$ and $3{d}_{3{z}^{2}-{r}^{2}}$ spectral weight, with $n \left( 3{d}_{{x}^{2}-{y}^{2}} \right) > n \left( 3{d}_{3{z}^{2}-{r}^{2}} \right)$ in general. Our practical goal is to extract the ratio ${f}_{zz} / {f}_{xx} $ at a photon energy $\omega \!=\! 931.5$\,eV, where all RXS measurements were performed. This is experimentally performed by measuring the XAS signal in the two geometries ${\boldsymbol \epsilon} \!\parallel\! \mathbf{x}$ and ${\boldsymbol \epsilon} \!\parallel\! \mathbf{z}$, which can be done by rotating the sample about the axis perpendicular to the scattering plane (see insets of Fig.\,S\ref{XAS}).

The experimental results at the Cu-${L}_{3}$ absorption edge on Bi2201 are shown in Fig.\,S\ref{XAS}a,b for the case ${\boldsymbol \epsilon} \!\parallel\! \mathbf{x}$ and ${\boldsymbol \epsilon} \!\parallel\! \mathbf{z}$, respectively. The $3{d}^{10}$ and $3{d}^{10}\underline{L}$ contributions can be separated and are best fitted using a Lorentzian and Gaussian peak, respectively, convoluted with a Gaussian resolution with 100\,meV spectral width. The energy positions and linewidths of the respective peaks are assumed to be independent of polarization, whereas the ratio between the transition strengths at the Cu-$2{p}^{6} 3{d}^{9}$ to Cu-$2{p}^{5} 3{d}^{10}$ features (blue peaks) provides an estimate of ${f}_{zz} / {f}_{xx} \!\simeq\! 0.15$. In the case of YBCO, a rough estimate can be done by looking at the undoped compound YBa$_{2}$Cu$_3$O$_{6}$ (where chains contribute no $3{d}^{10}$ final states), which yields a lower value of ${f}_{zz} / {f}_{xx} \!\sim\! 0.1$ [6]. Therefore we find a realistic experimental range to be $ 0.1 < {f}_{zz} / {f}_{xx} < 0.15$.
\\

\noindent {\bf From charge order symmetry to RXS model.} Intuitively, the on-site energies of the Cu-$2p$ and $3d$ orbitals are affected by the presence of extra charges located on the neighboring O-$2p$ states. In Refs. 7 and 8, RXS measurements of LNSCO and YBCO illustrated that the RXS transition energies were also spatially modulated in the presence of charge order in the CuO${}_{2}$ planes. Here we are interested in deriving a model incorporating the effect of a modulated charge distribution with different symmetry on the scattering tensor and, ultimately, on the measured RXS intensities. In the following, we lay out the general framework linking the energy shifts to the local charge modulations. However, we point out that, later on as we develop our RXS model, we will not rely on the detailed values of the energy shifts, but rather make use of the symmetry relation between the local form factor and the charge distribution under different symmetry configurations ($s$-, $s'$-, and $d$-wave).

In the absence of charge order, the valence charge is homogeneously distributed, so that each Oxygen site hosts an exact charge (in the $2p$ shell) of $ q \!=\! 6-p/2$, where $p$ is the nominal hole doping per Cu site. In the charge-ordered state, the electronic density forms a modulated pattern, $\Delta \rho \left( \mathbf{r} \right)$, with a maximum amplitude of ${\Delta \rho}_{\mathrm{max}}\!=\!{\Delta Q} / {V}_{\mathrm{UC}}$ (${V}_{\mathrm{UC}}$ is the unit cell volume). The spatially modulated charge produces a net crystal field (CF) acting on the Cu orbitals at site $\mathbf{R}$ through the Coulomb interaction [9]: 
\begin{equation}
{\Delta}_{i}^{\mathrm{CF}} (\mathbf{R}) = \frac{e}{{\varepsilon}_{\mathrm{eff}}} \iint \mathrm{d}\mathbf{r} \: \mathrm{d}\mathbf{{r}^{\prime}} \frac{\Delta \rho \left( \mathbf{r} \right) {\left\vert {\psi}_{i}^{\mathrm{Cu}} \left( \mathbf{{r}^{\prime}} - \mathbf{R} \right) \right\vert}^{2}}{\left\vert \mathbf{r}- \left( \mathbf{{r}^{\prime}} - \mathbf{R} \right) \right\vert},
\label{CF_def}
\end{equation}
\noindent
where ${\varepsilon}_{\mathrm{eff}}$ is an effective dielectric constant that accounts for the screening of the bare Coulomb potential, and ${\psi}_{i}^{\mathrm{Cu}}$ is the wavefunction of the $i$-th local Cu orbital, $i\!=\! \left\lbrace 2{p}_{x},2{p}_{y},2{p}_{z},3{d} \right\rbrace$. The density modulations can be decomposed into a site- (charges on Cu sites) and a bond- (charges on Oxygen sites) centered contribution:
\begin{equation}
\Delta {\rho}_{\mathrm{site}} \left( \mathbf{r}-{\mathbf{R}}_{i}\right) \!=\! \frac{\Delta {Q}}{2 {V}_{\mathrm{UC}}} \: \Delta_{\mathrm{CDW}} ({\mathbf{R}}_{i},{\mathbf{R}}_{i}){\left\vert {\psi}_{3{d}_{{x}^{2}-{y}^{2}}} (\mathbf{r} - {\mathbf{R}}_{i}) \right\vert}^{2}, \label{Delta_site}
\end{equation}
\begin{equation}
\Delta {\rho}_{\mathrm{bond}} \left( \mathbf{r}-{\mathbf{R}}_{i}\right) \!=\! \frac{\Delta {Q}}{2 {V}_{\mathrm{UC}}} \: \sum_{\langle j \rangle} \Delta_{\mathrm{CDW}} ({\mathbf{R}}_{i},{\mathbf{R}}_{j}) {\left\vert {\psi}_{2 {p}_{j}}\! \left( \mathbf{r}\!-\!\frac{{\mathbf{R}}_{i}+{\mathbf{R}}_{j}}{2} \right) \right\vert}^{2}\!,
\label{Delta_bond}
\end{equation}
\noindent
where in the second line the summation is over nearest-neighbor sites $\langle j \rangle$, and ${\psi}_{2 {p}_{j}}$ represents an O-$2 {p}_{x}$ or $2 {p}_{y}$ orbital depending on whether ${\mathbf{R}}_{i}+{\mathbf{R}}_{j}$ points along \textbf{x} or \textbf{y} (we only consider bonding O-$2p$ orbitals). Eqs.\,\ref{Delta_site} and \ref{Delta_bond} formalize the link between the density modulations and $\Delta_{\mathrm{CDW}} ({\mathbf{R}}_{i},{\mathbf{R}}_{j})$, namely the CDW order defined in real-space, which is related to its Fourier counterpart $\Delta_{\mathrm{CDW}} (\mathbf{k},\mathbf{Q}) \!=\! \left\langle {c}^{\dagger}_{\mathbf{k+Q/2}} \cdot {c}_{\mathbf{k-Q/2}} \right \rangle$ by [10]: 
\begin{equation}
\Delta_{\mathrm{CDW}} ({\mathbf{R}}_{i},{\mathbf{R}}_{j}) \propto \sum_{\mathbf{Q}} \sum_{\mathbf{k}} \Delta_{\mathrm{CDW}} (\mathbf{k},\mathbf{Q}) {e}^{i \mathbf{k} {\textstyle\cdot} \: ({\mathbf{R}}_{i}-{\mathbf{R}}_{j})} {e}^{i \mathbf{Q} {\textstyle\cdot} \: ({\mathbf{R}}_{i}+{\mathbf{R}}_{j})/2}.
\label{Delta_CDW_def}
\end{equation}
\noindent
At this point, $\Delta_{\mathrm{CDW}} (\mathbf{k}, \mathbf{Q})$ can be expanded as follows [10]:
%
%
\begin{equation}
\Delta_{\mathrm{CDW}} (\mathbf{k},\pm {\mathbf{Q}}_{\mathrm{CDW}}) = \left[ {\Delta}_{s}+{\Delta}_{{s}^{\prime}} (\cos {k}_{x}+\cos {k}_{y})+{\Delta}_{d} (\cos {k}_{x}-\cos {k}_{y}) \right]
 \label{Delta_CDW_k_Q}
\end{equation}
where ${\Delta}_{s}$ is the representation for site-centered CDW (with \textit{s}-wave symmetry), while ${\Delta}_{{s}^{\prime}}$ and ${\Delta}_{d}$ are associated with an extended \textit{s}- and a \textit{d}-wave bond order, respectively. The three terms in Eq.\,\ref{Delta_CDW_k_Q} are treated independently in subsequent calculations.

Using Eqs.\,\ref{Delta_site}, \ref{Delta_bond}, and \ref{Delta_CDW_def}, we can express the charge modulations (at site $n$) for the Cu-$3d$ and O-$2p$ orbitals associated to each symmetry term in Eq.\,\ref{Delta_CDW_k_Q} as follows:
\begin{eqnarray}
\mbox{\textit{s}-}\!\!\!\!\! &\mbox{wave}& \begin{cases} {\Delta \rho}_{\mathrm{site}} \left( \mathrm{Cu} \right) \propto \cos \left( {\mathbf{Q}}_{\mathrm{CDW}} \cdot {\mathbf{R}}_{n} \right) \end{cases}\nonumber \\
\mbox{\textit{s'}-}\!\!\!\!\! &\mbox{wave}& \begin{cases} {\Delta \rho}_{\mathrm{bond}} \left( {\mathrm{O}}_{x \pm} \right) \propto \cos \left( {\mathbf{Q}}_{\mathrm{CDW}} \cdot \left( {\mathbf{R}}_{n} \pm a/2 \: \mathbf{\hat{x}} \right) \right) \\
{\Delta \rho}_{\mathrm{bond}} \left( {\mathrm{O}}_{y \pm} \right) \propto \cos \left( {\mathbf{Q}}_{\mathrm{CDW}} \cdot {\mathbf{R}}_{n} \right) \end{cases}\nonumber \\
\mbox{\textit{d}-}\!\!\!\!\! &\mbox{wave}& \begin{cases} {\Delta \rho}_{\mathrm{bond}} \left( {\mathrm{O}}_{x \pm} \right) \propto \cos \left( {\mathbf{Q}}_{\mathrm{CDW}} \cdot \left( {\mathbf{R}}_{n} \pm a/2 \: \mathbf{\hat{x}} \right) \right) \\
{\Delta \rho}_{\mathrm{bond}} \left( {\mathrm{O}}_{y \pm} \right) \propto \cos \left( {\mathbf{Q}}_{\mathrm{CDW}} \cdot {\mathbf{R}}_{n} + \pi \right),
\end{cases}
\label{DeltaQ_symmetry}
\end{eqnarray}
where ${\mathrm{O}}_{x \pm}$ and ${\mathrm{O}}_{y \pm}$ represent the O-$2p$ orbitals located at $ {\mathbf{R}}_{n} \pm a/2 \: \mathbf{\hat{x}} $ and $ {\mathbf{R}}_{n} \pm a/2 \: \mathbf{\hat{y}} $, respectively.

The RXS signal arises because the form factor is spatially modulated about an average value, $ {f}_{pq} \!=\! \left[ {\bar{f}}_{pq} + {\Delta f}_{pq} \right] {\delta}_{pq}^{n} $. The scattering tensor can then be explicitly calculated using:
\begin{eqnarray}
{F}_{pq} \left( \mathbf{Q} \right) &=& \dfrac{1}{N} {\sum}_{n} \left\vert \begin{array}{ccc} {\bar{f}}_{xx} + {\Delta f}_{xx}^{n} & 0 & 0 \\ 0 & {\bar{f}}_{yy} + {\Delta f}_{yy}^{n} & 0 \\ 0 & 0 & {\bar{f}}_{zz} + {\Delta f}_{zz}^{n} \end{array} \right\vert {e}^{i \mathbf{Q} \cdot {\mathbf{R}}_{n}} \nonumber \\
&=& \dfrac{1}{N} {\sum}_{n}  \left\vert \begin{array}{ccc} {\bar{f}}_{xx} & 0 & 0 \\ 0 & {\bar{f}}_{yy} & 0 \\ 0 & 0 & {\bar{f}}_{zz}\end{array} \right\vert {e}^{i \mathbf{Q} \cdot {\mathbf{R}}_{n}} + \left\vert \begin{array}{ccc} {\Delta f}_{xx}^{n} & 0 & 0 \\ 0 & {\Delta f}_{yy}^{n} & 0 \\ 0 & 0 & {\Delta f}_{zz}^{n} \end{array} \right\vert {e}^{i \mathbf{Q} \cdot {\mathbf{R}}_{n}} \nonumber \\
&=& \dfrac{1}{N} {\sum}_{n}  \left\vert \begin{array}{ccc} {\Delta f}_{xx}^{n} & 0 & 0 \\ 0 & {\Delta f}_{yy}^{n} & 0 \\ 0 & 0 & {\Delta f}_{zz}^{n} \end{array} \right\vert {e}^{i \mathbf{Q} \cdot {\mathbf{R}}_{n}} 
\label{Scattering_factor_vs_form_factor}
\end{eqnarray}
where the last line assumes that $\mathbf{Q} \!\neq\! 0$, a condition which causes the first term in the second line to vanish. It is clear from Eq.\,\ref{Scattering_factor_vs_form_factor} that spatial variations of ${f}_{pq}$ are an essential ingredient to have a nonzero scattering tensor and, therefore, a nonzero RXS cross section. The variations in the local form factor ${f}_{pq}$ considered in our model are a consequence of a spatial variation in the energy shifts $\Delta E$, which in turn are determined by the fluctuations in the local electronic density ($\Delta \rho$), according to Eq.\ref{CF_def}. For small amplitudes of $\Delta \rho$ (typically in cuprates the charge inhomogeneity is of the order of $\Delta \rho < 0.1e$ [11-12]) and consequently of $\Delta E$, we can Taylor-expand $\Delta f$ with respect to $\Delta \rho$ and retain only the lowest (linear) order, which leads to:
\begin{eqnarray}
\mbox{\textit{s}-}\!\!\!\!\! &\mbox{wave}& \begin{cases} {\Delta f}_{xx}^{(n)} = {\Delta f}_{yy}^{(n)} \propto \Delta \rho \left( \mathrm{Cu} \right) \end{cases}\nonumber \\
\mbox{\textit{s'}- and \textit{d}-}\!\!\!\!\! &\mbox{wave}& \begin{cases} {\Delta f}_{xx}^{(n)} \propto \Delta \rho \left( {\mathrm{O}}_{x+} \right) + \Delta \rho \left( {\mathrm{O}}_{x-} \right) \\ {\Delta f}_{yy}^{(n)} \propto \Delta \rho \left( {\mathrm{O}}_{y+} \right) + \Delta \rho \left( {\mathrm{O}}_{y-} \right)  \end{cases}
\label{Deltaf_expansion}
\end{eqnarray}
Where the expansions take into account the fact that the core-to-valence transitions under considerations are more sensitive to local variations in the occupation of certain orbitals, e.g. in presence of \textit{s'}- or \textit{d}-wave order the ${\Delta f}_{xx}^{(n)}$ (${\Delta f}_{yy}^{(n)}$) terms are primarily sensitive to variations in the density of the x- (y-) coordinated O-$2p$ orbitals, since they reflect transitions involving initial states that are pointing in the x (y) direction, i.e. Cu-${2p}_{x}$ (Cu-${2p}_{y}$).

Combining Eq.\,\ref{DeltaQ_symmetry} and \ref{Deltaf_expansion} leads to the following core expression for our RXS model:
\begin{eqnarray}
\mbox{\textit{s}-}\!\!\!\!\! &\mbox{wave}& \begin{cases} {\Delta f}_{xx}^{(n)} &\!\!\!\!\!= {\Delta f}_{yy}^{(n)} = {\delta}_{s} \cos \left( {\mathbf{Q}}_{\mathrm{CDW}} \cdot {\mathbf{R}}_{n} \right) \\ {\Delta f}_{zz}^{(n)} &\!\!\!\!\!= \gamma \times {\delta}_{s} \cos \left( {\mathbf{Q}}_{\mathrm{CDW}} \cdot {\mathbf{R}}_{n} \right) \end{cases}\nonumber \\
\mbox{\textit{s'}-}\!\!\!\!\! &\mbox{wave}& \begin{cases} {\Delta f}_{xx}^{(n)} &\!\!\!\!\!= \frac{1}{2} \times {\delta}_{s'} \left[ \cos \left( {\mathbf{Q}}_{\mathrm{CDW}} \cdot \left( {\mathbf{R}}_{n} + a/2 \: \mathbf{\hat{x}} \right) \right) + \cos \left( {\mathbf{Q}}_{\mathrm{CDW}} \cdot \left( {\mathbf{R}}_{n} - a/2 \: \mathbf{\hat{x}} \right) \right) \right]  \\
&\!\!\!\!\!= {\delta}_{s'} \cdot \cos \left( {\mathbf{Q}}_{\mathrm{CDW}} \cdot {\mathbf{R}}_{n} \right) \cdot \cos \phi \\
{\Delta f}_{yy}^{(n)} &\!\!\!\!\!= {\delta}_{s'} \cos \left( {\mathbf{Q}}_{\mathrm{CDW}} \cdot {\mathbf{R}}_{n} \right) \\ {\Delta f}_{zz}^{(n)} &\!\!\!\!\!= 0 \end{cases}\nonumber \\
\mbox{\textit{d}-}\!\!\!\!\! &\mbox{wave}& \begin{cases} {\Delta f}_{xx}^{(n)} &\!\!\!\!\!= \frac{1}{2} \times {\delta}_{d} \left[ \cos \left( {\mathbf{Q}}_{\mathrm{CDW}} \cdot \left( {\mathbf{R}}_{n} + a/2 \: \mathbf{\hat{x}} \right) \right) + \cos \left( {\mathbf{Q}}_{\mathrm{CDW}} \cdot \left( {\mathbf{R}}_{n} - a/2 \: \mathbf{\hat{x}} \right) \right) \right]  \\
&\!\!\!\!\!= {\delta}_{d} \cdot \cos \left( {\mathbf{Q}}_{\mathrm{CDW}} \cdot {\mathbf{R}}_{n} \right) \cdot \cos \phi \\
{\Delta f}_{yy}^{(n)} &\!\!\!\!\!= {\delta}_{d} \cos \left( {\mathbf{Q}}_{\mathrm{CDW}} \cdot {\mathbf{R}}_{n} + \pi \right) = - {\delta}_{d} \cos \left( {\mathbf{Q}}_{\mathrm{CDW}} \cdot {\mathbf{R}}_{n} \right) \\ {\Delta f}_{zz}^{(n)} &\!\!\!\!\!= 0, \end{cases}
\label{Deltaf_master_equations}
\end{eqnarray}
where we have introduced the phase $\phi \!=\! {\mathbf{Q}}_{\mathrm{CDW}} \cdot a/2 \: \mathbf{\hat{x}} $ and set $\gamma \!=\! {\Delta f}_{zz} /  {\Delta f}_{xx} \!=\! {\bar{f}}_{zz} /  {\bar{f}}_{xx}$ to represent the anisotropy ratio in the form factor tensor. The magnitudes of the \textit{s}-, \textit{s'}-, and \textit{d}-wave components of the charge order are here indicated as ${\delta}_{s}$, ${\delta}_{s'}$, and ${\delta}_{d}$, respectively. Note that out-of-plane transition at the Cu site are hardly affected by small variations in the O 2p charge due to the small intersite orbital overlap, hence we have set ${\Delta f}_{zz}^{(n)} \!=\! 0$ for the \textit{s'}- and \textit{d}-wave case. This parametrization and subsequent expression in terms of CDW symmetry is similar to one developed in Ref. 2, which considered it in the case of scattering from the O sublattice.

The above expressions can now be inserted into Eq.\,\ref{Scattering_factor_vs_form_factor} to derive the scattering tensor:
\begin{eqnarray}
\mbox{\textit{s}-}\!\!\!\!\! &\mbox{wave}& \!\!\!\! \begin{cases} {F}_{pq} \left( {\mathbf{Q}}_{\mathrm{CDW}} \right) = \dfrac{1}{N} {\sum}_{n} {\delta}_{s} \cos \left( {\mathbf{Q}}_{\mathrm{CDW}} \cdot {\mathbf{R}}_{n} \right) \left\vert \begin{array}{ccc} 1 & 0 & 0 \\ 0 & 1 & 0 \\ 0 & 0 & \gamma \end{array} \right\vert {e}^{i {\mathbf{Q}}_{\mathrm{CDW}} \cdot {\mathbf{R}}_{n}} = {\delta}_{s} {F}_{pq}^{(s)} \end{cases}\nonumber \\
\mbox{\textit{s'}-}\!\!\!\!\! &\mbox{wave}& \!\!\!\! \begin{cases} {F}_{pq} \left( {\mathbf{Q}}_{\mathrm{CDW}} \right) = \dfrac{1}{N} {\sum}_{n} {\delta}_{s'} \cos \left( {\mathbf{Q}}_{\mathrm{CDW}} \cdot {\mathbf{R}}_{n} \right) \left\vert \begin{array}{ccc} \cos \phi & 0 & 0 \\ 0 & 1 & 0 \\ 0 & 0 & 0 \end{array} \right\vert {e}^{i {\mathbf{Q}}_{\mathrm{CDW}} \cdot {\mathbf{R}}_{n}} = {\delta}_{s'} {F}_{pq}^{(s')} \end{cases}\nonumber \\
\mbox{\textit{d}-}\!\!\!\!\! &\mbox{wave}& \!\!\!\! \begin{cases} {F}_{pq} \left( {\mathbf{Q}}_{\mathrm{CDW}} \right) = \dfrac{1}{N} {\sum}_{n} {\delta}_{d} \cos \left( {\mathbf{Q}}_{\mathrm{CDW}} \cdot {\mathbf{R}}_{n} \right) \left\vert \begin{array}{ccc} \cos \phi & 0 & 0 \\ 0 & -1 & 0 \\ 0 & 0 & 0 \end{array} \right\vert {e}^{i {\mathbf{Q}}_{\mathrm{CDW}} \cdot {\mathbf{R}}_{n}} = {\delta}_{d} {F}_{pq}^{(d)} \end{cases}
\label{eq:Scattering_tensor_final}
\end{eqnarray}

Using these equations we can write, in a more compact form, the scattering tensor associated to a linear combination of ${\delta}_{s}$, ${\delta}_{s'}$, and ${\delta}_{d}$:
\begin{equation}
{F}_{pq} \left( \pm {\mathbf{Q}}_{\mathrm{CDW}} \right) = \left\vert \begin{array}{ccc} {\delta}_{s} + \left( {\delta}_{s'} + {\delta}_{d} \right) \cos \phi & 0 & 0 \\ 0 & {\delta}_{s} + {\delta}_{s'} - {\delta}_{d} & 0 \\ 0 & 0 & \gamma {\delta}_{s} \end{array} \right\vert
\label{Scattering_tensor_combination}
\end{equation}
This last expression -- depending exclusively on the magnitudes of the \textit{s}-, \textit{s'}-, and \textit{d}-wave symmetry terms, on the wavevector- (and therefore sample-) dependent phase $\phi \!=\! {\mathbf{Q}}_{\mathrm{CDW}} \cdot a/2 \: \mathbf{\hat{x}} $, and on the parameter $\gamma$ -- has been ultimately used to model the azimuthal-dependent RXS signal as explained in more detail in the next section. We note that the use of a model that is based on the general symmetry of the charge distribution (rather than on the microscopic charge pattern), such as the one which is condensed in Eq.\,\ref{eq:Scattering_tensor_final} implies that our framework cannot in principle be used to distinguish between a stripe-like and a checkerboard-like scenario.\\

\noindent {\bf RXS azimuthal angle dependence and validity of charge-ordering models.} Calculations of the scattering intensity as measured using RXS have been performed starting from Eq.\,\ref{RXS_def}, and using the functional form for the scattering tensor as given in Eq.\,\ref{Scattering_tensor_combination}. For each azimuthal angle $\alpha$ the scattering tensor $\mathbf{F}$ is transformed using the rotation matrix ${\mathbf{R}}_{\hat{\mathbf{u}}} (\alpha)$ ($\hat{\mathbf{u}}$ is the azimuthal rotation axis, which is parallel to the wavevector $\mathbf{Q}$), yielding $\mathbf{\tilde{F}} (\alpha)\!=\! {\mathbf{R}}_{\hat{\mathbf{u}}} (\alpha) \cdot \mathbf{F} \cdot {\mathbf{R}}_{\hat{\mathbf{u}}}^{\top} (\alpha)$. Replacing this last expression in the formula for the RXS cross section (Eq.\,\ref{RXS_def}) leads to the master expression used to generate the theoretical RXS azimuthal profiles:
\begin{equation}
{I}^{\mathrm{RXS}} (\alpha) = {\left\vert {\sum}_{p q} {\epsilon}_{p} \cdot {\tilde{F}}_{pq} (\alpha) \cdot {\epsilon}^{\prime}_{q} \right\vert}^{2},
\label{Scattering_tensor_rotation}
\end{equation}
Note that in our model the polarization vectors are assumed to be fixed (since they belong to the laboratory frame of reference). The calculated profiles are subsequently corrected for self-absorption using the formula [13]: 
\begin{equation}
{I}_{\mathrm{calc}} (\alpha) \!=\!{I}^{\mathrm{RXS}} (\alpha) \times\! {\left[ {\mu}_{in} + {\mu}_{out} \times \frac{\cos \left( {\mathbf{k}}_{in} (\alpha) \cdot \hat{\mathbf{n}} (\alpha) \right)}{\cos \left( {\mathbf{k}}_{out} (\alpha) \cdot \hat{\mathbf{n}} (\alpha) \right)} \right]}^{-1},
\label{SA_correction}
\end{equation}
where ${\mathbf{k}}_{in,out}$ represent the incident and scattered wavevectors, respectively, while $\hat{\mathbf{n}}$ is the surface normal. The projections of the absorption tensor ${\mu}_{ij}$ onto the incoming and outgoing x-ray polarizations are denoted as ${\mu}_{in} \!=\! {\epsilon}_{i} {\mu}_{ij} {\epsilon}_{j} $ and ${\mu}_{out} \!=\! {\epsilon}^{\prime}_{i} {\mu}_{ij} {\epsilon}^{\prime}_{j} $, respectively; the absorption tensor in cuprates is diagonal with ${\mu}_{xx} \!=\! {\mu}_{yy}$ and ${\mu}_{zz} \!\simeq\! 0.6 \times {\mu}_{xx}$ [6]. In order to compare the theory and the measurements on equal grounds, there are two options: (i) to remove the self-absorption contribution from the experimental data; or (ii) to incorporate the self-absorption correction into the numerical calculations. Since the self-absorption correction depends on both the incoming ($\boldsymbol \epsilon$) and outgoing ($\boldsymbol {\epsilon}^{\prime}$) polarization vectors (through the absorption tensor ${\mu}_{ij}$), applying the correction directly onto the experimental data [case (i)] is not applicable, since the amount of light scattered in each outgoing polarization channel was not resolved in the experiments. Therefore we have applied the self-absorption correction onto the calculated profiles, where instead we have full knowledge of the polarization vectors that enter Eq.\,\ref{Scattering_tensor_rotation}.
\begin{figure}[t!]
\centering
\includegraphics[width=1\linewidth]{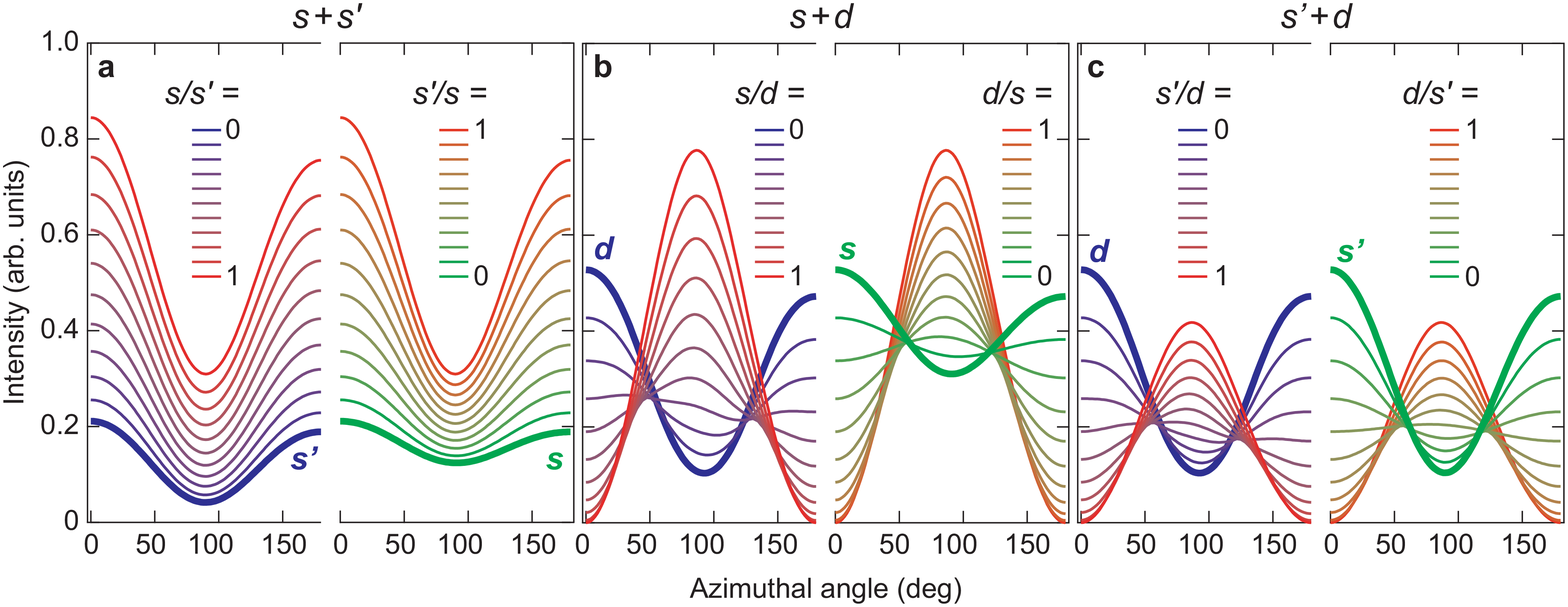}
\caption{{\bf Calculated azimuthal profiles for hybrid combinations of charge order symmetry terms.} Calculated profiles ${I}_{\mathrm{calc}} (\alpha)$, using Eqs.\,\ref{Scattering_tensor_rotation} and \ref{SA_correction}, for mixed orders combinations: \textbf{a}, $s + {s}^{\prime}$; \textbf{b}, $s + d$; \textbf{c}, ${s}^{\prime} + d$. Thicker lines represent the single-symmetry terms $s$, ${s}^{\prime}$, and $d$.}
\label{Mix_order_azimuthal_profiles}
\end{figure}

The ${I}_{\mathrm{calc}} (\alpha)$ profiles have been calculated for a linear combination of the symmetry terms of the CDW order $\Delta_{\mathrm{CDW}} (\mathbf{k},\mathbf{Q})$, as encoded in Eq.\ref{Scattering_tensor_combination}. A value of $\gamma \!=\! 0.1$ ($\gamma \!=\! 0.15$) has been used for YBCO (Bi2201). The expression in Eq.\ref{Scattering_tensor_combination} has been treated as a model function with fitting parameters ${\delta}_{s}$, ${\delta}_{s'}$, and ${\delta}_{d}$, and a chi-square minimization with respect to the measured datasets has been performed. However, if all three symmetry components are assumed to be nonzero and free to vary, the fitting procedures are found to inevitably converge to local minima which depend on the choice of initial guesses. This is in part due to the fact that the scattering tensor, despite having three nonzero entries (${F}_{xx}$, ${F}_{yy}$, and ${F}_{zz}$), can be uniquely identified by an irreducible set of only two parameters (e.g., ${F}_{zz} / {F}_{xx}$ and ${F}_{yy} / {F}_{xx}$), since an overall rescaling of the tensor will simply yield an amplitude rescaling. As a direct consequence of this fact, any attempt to fit the experimental data with three free parameters leads to large cross-correlations in the fitting coefficients and to a failure of the nonlinear regression procedure.
Therefore, our fitting analysis has been constrained to a combination of at most \textit{two} symmetry terms, i.e. for the six possible cases: (i) $s$; (ii) ${s}^{\prime}$; (iii) $d$; (iv) $s + {s}^{\prime}$; (v) $s + d$; and (vi) ${s}^{\prime} + d$. In any case, as will become clear later, the data are already well-reproduced with a combination of two symmetry terms, lifting the need for a 3-component fit. The azimuthal profiles arising from combinations of this kind, for the case of vertical ($\sigma$) incoming polarization, are shown in Fig.\,S\ref{Mix_order_azimuthal_profiles} (the single symmetry profiles are represented by thicker lines). Furthermore, since the scattering tensor does not depend on the incoming light polarization $\boldsymbol \epsilon$ (the charge order symmetry is an intrinsic property of the system and therefore does not change with the probing geometry), the datasets acquired using vertical and horizontal light polarizations (for the same compound) have been fitted with the added constraint that all the fitting parameters be the same for the two polarizations.

\begin{figure}[t!]
\centering
\includegraphics[width=0.85\linewidth]{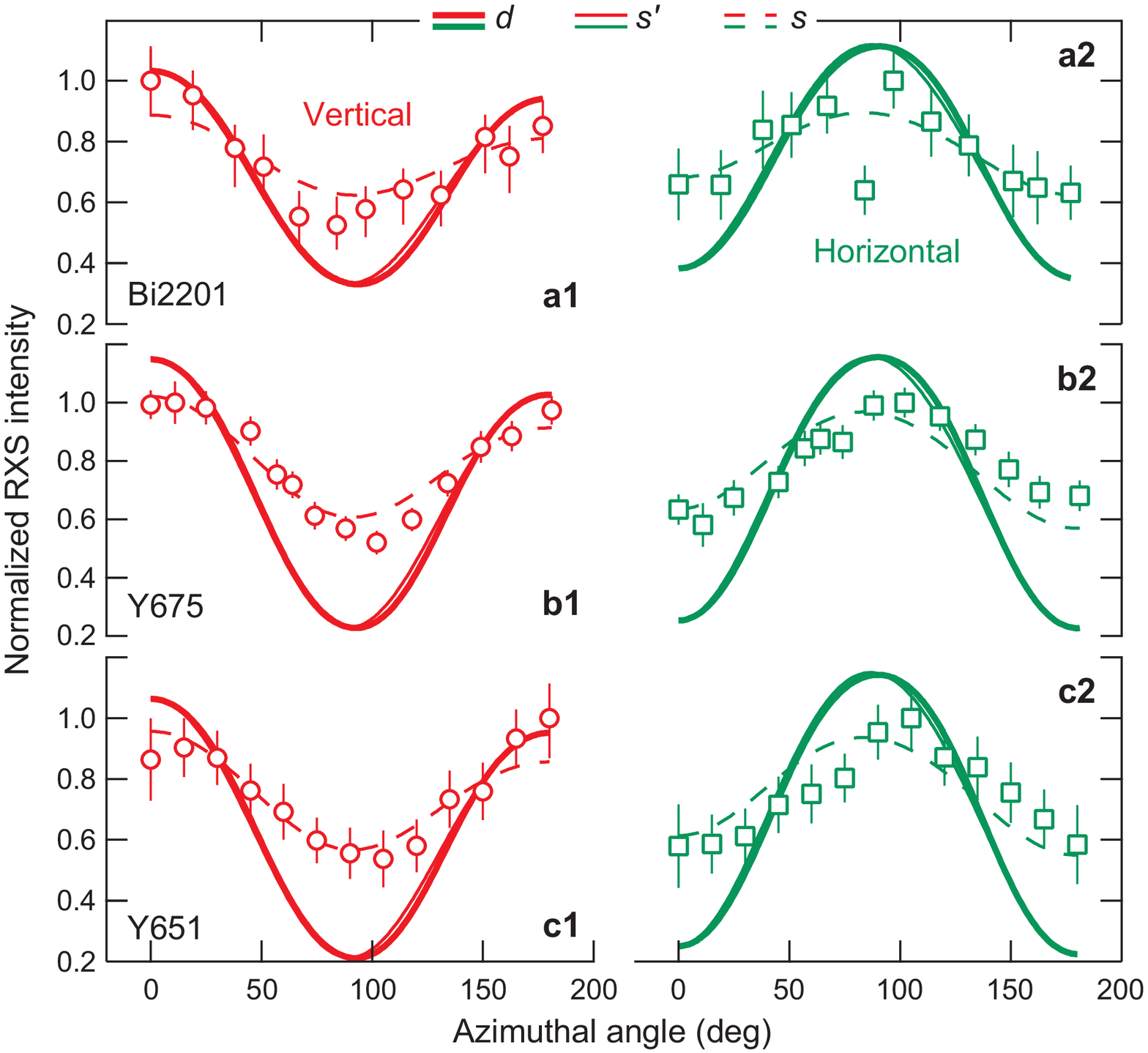}
\caption{{\bf Experimental and calculated (single-symmetry) CDW peak intensity vs. azimuthal angle.} Normalized RXS intensities (uncorrected for self-absorption) are plotted using red (green) markers for scans acquired using vertical (horizontal) incoming polarization. Theoretical profiles (corrected for self-absorption) for three possible combinations are obtained from a least-squares fitting method and overlaid on the data -- $d$ (full thick), ${s}^{\prime}$ (full thin), and $s$ (dashed thin). Datasets are presented for: \textbf{a1,a2}, Bi2201; \textbf{b1,b2}, Y675; \textbf{c1,c2}, Y651.}
\label{All_azimuthal_profiles_single_symmetry}
\end{figure}
\begin{figure}[t!]
\centering
\includegraphics[width=0.85\linewidth]{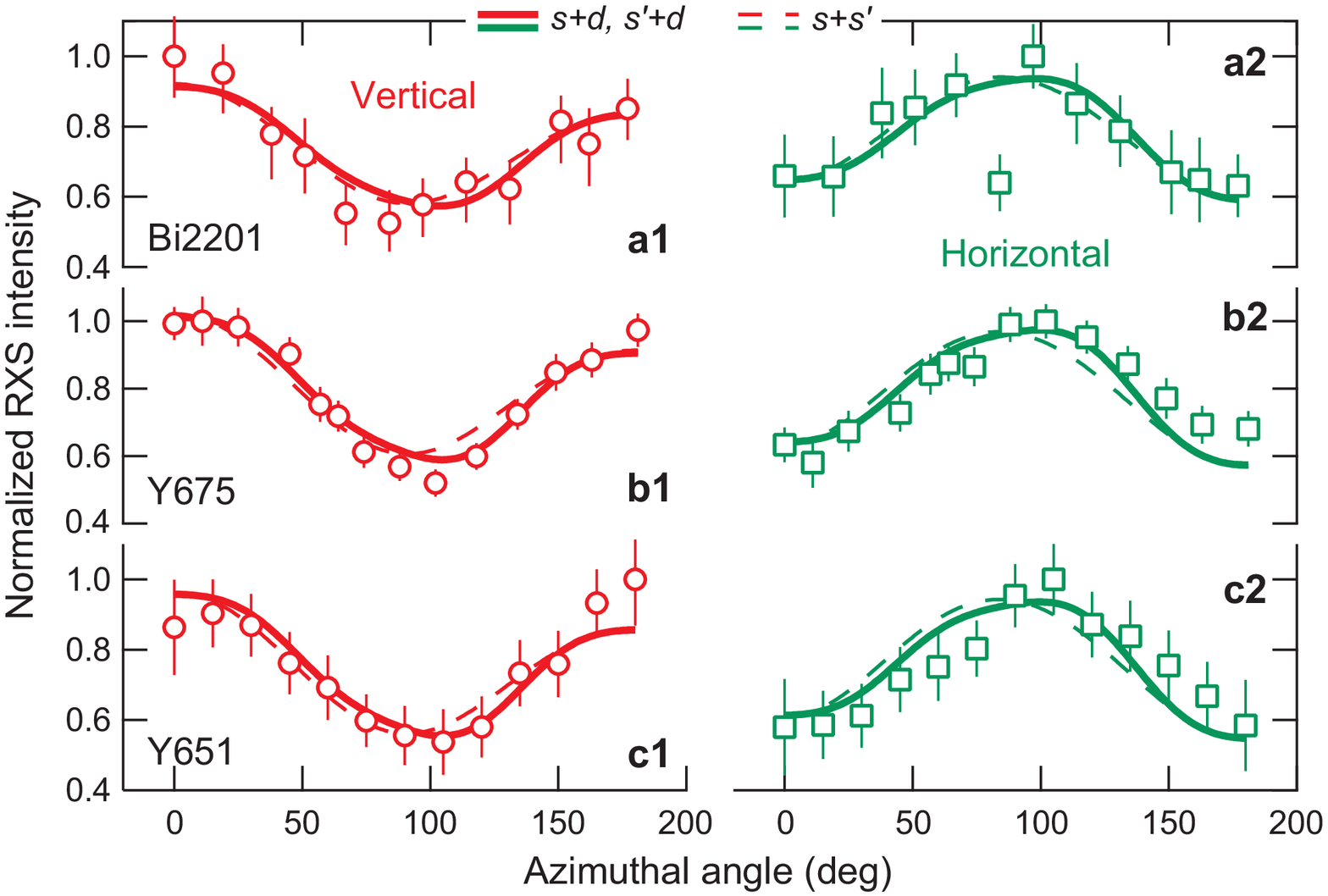}
\caption{{\bf Experimental and calculated (mixed-symmetry) CDW peak intensity vs. azimuthal angle.} Normalized RXS intensities (uncorrected for self-absorption) are plotted using red (green) markers for scans acquired using vertical (horizontal) incoming polarization. Theoretical profiles (corrected for self-absorption) for three possible combinations are obtained from a least-squares fitting method and overlaid to the data: $s + d$ and ${s}^{\prime} + d$ (full, a single trace is used since the resulting profiles are nearly overlapping), and $s + {s}^{\prime}$ (dashed). Datasets are presented for: \textbf{a1,a2}, Bi2201; \textbf{b1,b2}, Y675; \textbf{c1,c2}, Y651.}
\label{All_azimuthal_profiles_mixed_symmetry}
\end{figure}

The experimental RXS azimuthal intensities for both vertical and horizontal incoming polarization are reported in Fig.\,S\ref{All_azimuthal_profiles_single_symmetry} and S\ref{All_azimuthal_profiles_mixed_symmetry}, with error bars obtained from a non-linear least-squares regression analysis using a gradient-based method to determine the best fit parameters. Figure Fig.\,S\ref{All_azimuthal_profiles_single_symmetry} shows the fit results for single-symmetry terms, which would at first glance suggest that the $s$-wave model is the one which best reproduces the experimental data. However, the use of a combination of two symmetry terms, shown in Fig.\,S\ref{All_azimuthal_profiles_mixed_symmetry}, reveals how the addition of a second component brings all of the best-fit theoretical profiles closer together. In particular, the mixed terms containing a $d$-wave term are found to interpolate the data more closely than the purely symmetric combination $s + {s}^{\prime}$.
This is explained by the observation that the symmetric \textit{s}- and \textit{s'}-wave terms and any linear combination fail at reproducing the experimental data since they always yield a symmetric distribution of intensities, centered about the azimuthal angle $\alpha \!=\! {90}^{\circ}$, in contrast to the experimental data from YBCO which showcase a $\sim {10}^{\circ}$ shift of min/max away from ${90}^{\circ}$.

\begin{figure}[t!]
\centering
\includegraphics[width=1\linewidth]{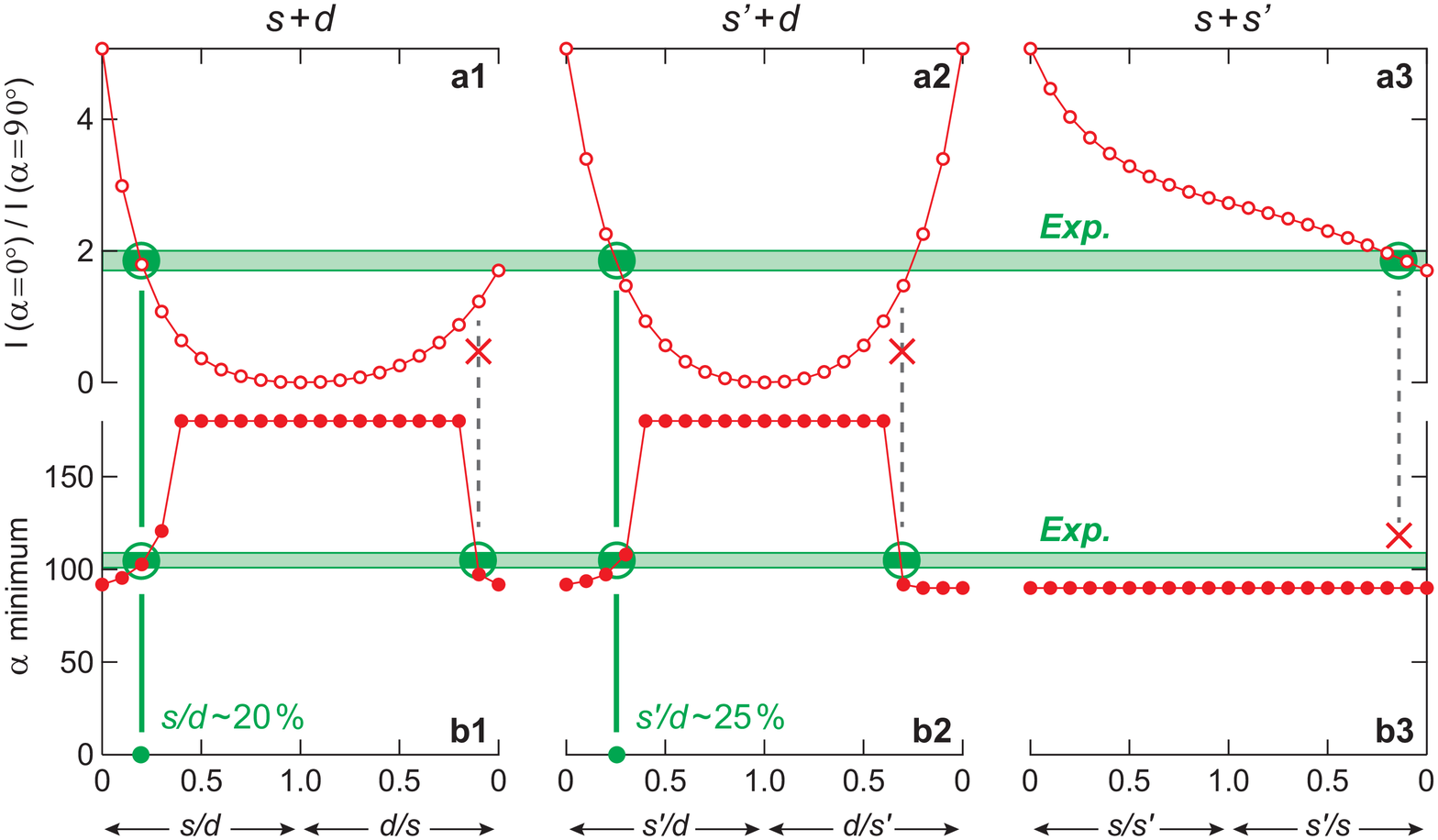}
\caption{{\bf Intensity ratio and minimum vs. azimuthal angle for the three binary models.} \textbf{a1,a2,a3}, Ratio of the calculated RXS intensities at $\alpha \!=\!{0}^{\circ}$ and $\alpha \!=\!{90}^{\circ}$ (assuming vertical polarization) as a function of the symmetry term mixing ratio for $s+d$, ${s}^{\prime}+d$, and $s+{s}^{\prime}$, respectively. The shaded horizontal bar marks the experimental range for YBCO. \textbf{b1,b2,b3}, Angular position of the minimum RXS intensity (again assuming vertical polarization) as a function of the symmetry term mixing ratio for $s+d$, ${s}^{\prime}+d$, and $s+{s}^{\prime}$, respectively. The shaded horizontal bar marks the experimental range for YBCO.}
\label{Minimum_amplitude_exp_vs_model}
\end{figure}
This situation is analyzed in more detail by calculating the intensity ratio between $\alpha\!=\!{0}^{\circ}$ and $\alpha\!=\!{90}^{\circ}$ [${I}_{\mathrm{RXS}} \left( \alpha \!=\! {0}^{\circ} \right) / {I}_{\mathrm{RXS}} \left( \alpha \!=\! {90}^{\circ} \right)$] and the minimum angular position in the azimuthal RXS intensity distribution ${\alpha}_{\mathrm{min}}$, which serve as useful metrics for the assessment of the similarity between the calculated profiles and the experimental data. The corresponding traces, evaluated for the symmetry combination $s+d$, ${s}^{\prime}+d$, and $s+{s}^{\prime}$ (and assuming vertical light polarization) as a function of the symmetry term mixing ratios are shown in Fig.\,S\ref{Minimum_amplitude_exp_vs_model}: a1-a3 and b1-b3, respectively. The experimental bands are represented by the shaded area. The best agreement between theoretical and experimental RXS intensities can be estimated to occur when the calculated traces in Fig.\,S\ref{Minimum_amplitude_exp_vs_model} cross the experimental bands, for both ${I}_{\mathrm{RXS}} \left( \alpha \!=\! {0}^{\circ} \right) / {I}_{\mathrm{RXS}} \left( \alpha \!=\! {90}^{\circ} \right)$ and ${\alpha}_{\mathrm{min}}$. These simultaneous crossing points are marked with green circles, and can be seen to occur only in the  $s+d$ and ${s}^{\prime} + d$ cases.  The reason why such crossing is not present for a $s+{s}^{\prime}$ combination follows from the fact that the minimum RXS intensity is always found at $\alpha\!=\!{90}^{\circ}$ (see Fig.\,\ref{Minimum_amplitude_exp_vs_model}b3). On the other hand, for combinations containing a $d$-wave component, the theoretical curves are found to approach the experimental data when $s / d \!=\! {\delta}_{s} / {\delta}_{d} \!\sim\! 0.2$ (${s}^{\prime} / d \!=\! {\delta}_{{s}^{\prime}} / {\delta}_{d} \!\sim\! 0.25$), thus providing a qualitative picture for the results of the fitting analysis, which returned very consistent values. While there are other values of $s/d$ and ${s}^{\prime}/ d$ that verify one or the other condition (yielding a crossing for ${I}_{\mathrm{RXS}} \left( \alpha \!=\! {0}^{\circ} \right) / {I}_{\mathrm{RXS}} \left( \alpha \!=\! {90}^{\circ} \right)$ or ${\alpha}_{\mathrm{min}}$), it is only when $s / d \!\sim\! 0.2$  (${s}^{\prime} / d \!\sim\! 0.25$) that these two conditions are verified at the same time. Ultimately, and regardless of the specific model parameters assumed for the scattering tensor, it is the very structure of the latter, with a sign change in the \textit{d}-wave component in Eq.\,\ref{Scattering_tensor_combination}, that underlies the need for a \textit{d}-wave term to reproduce the slight skewness in the azimuthal modulation of the RXS intensities. 

In addition, we note that, since the proportionality between the scattering yield and the charge modulation amplitude (the latter being proportional to ${\Delta}_{\mathrm{CDW}}$) is not exactly the same for charges sitting on the Cu site (\textit{s}-wave order) and O site (\textit{s'}- and \textit{d}-wave order), we can write that ${\delta}_{d} / {\delta}_{{s}^{\prime}} \!=\! {\Delta}_{d} / {\Delta}_{{s}^{\prime}}$, while in general ${\delta}_{s} / {\delta}_{d} \! > \! {\Delta}_{s} / {\Delta}_{d}$ and ${\delta}_{s} / {\delta}_{{s}^{\prime}} \! > \! {\Delta}_{s} / {\Delta}_{{s}^{\prime}} $ (the last two inequalities following from the fact that the energy shifts are larger in presence of extra charges residing on the site rather than in the bond).

\begin{table}[b!]
\begin{center}
\begin{tabular*}{0.6\linewidth}{@{\extracolsep{\fill}}P{0.15\linewidth}P{0.15\linewidth}P{0.15\linewidth}P{0.15\linewidth}@{}}
\hline
\hline
& & & \\
 & \multicolumn{3}{c}{Cumulative reduced chi-square ${\chi}^{2}_{\mathrm{tot}}$} \\
& & & \\
 & $ s + {s}^{\prime} $ & $ s + d $ & $ {s}^{\prime} + d $ \\
\hline
Bi2201 & 0.95 & 0.96 & 0.96 \\
Y651 & 0.72 & 0.49 & 0.50 \\
Y675 & 1.94 & 1.03 & 1.05 \\
\hline
\hline
\end{tabular*}
\label{table_report}
\end{center}
\caption{{\bf Goodness-of-fit for various combinations of \textit{s}-, \textit{s'}-, and \textit{d}-wave symmetry components}. Values of the cumulative reduced chi-square ${\chi}^{2}_{\mathrm{tot}}$, obtained after fitting the entire dataset to the various combinations of symmetry terms under consideration: $ s + {s}^{\prime} $, $ s + d $, and $ {s}^{\prime} + d $.}
\label{Chi_square_analysis}
\end{table}

As a figure of merit to evaluate the validity of the models with respect to the experimental results, we have used the reduced chi-square ${\chi}^{2}_{\mathrm{red}}$ [14] defined as follows:
\begin{equation}
{\chi}^{2}_{\mathrm{red}} = \frac{1}{N\!-\!3} \sum_{p=1}^{N} {\left( \frac{{I}_{p} - {I}_{\mathrm{calc}} ({\alpha}_{p})}{{\sigma}_{p}} \right)}^{2},
\end{equation}
\noindent
where ${I}_{p}$ and ${\alpha}_{p}$ are the experimental data (RXS intensities and azimuthal angles, respectively, from the data shown in Fig.\,4 in the main text); ${\sigma}_{p}$ are the uncertainties in the determination of the scattering intensities ${I}_{p}$, derived from Gaussian fits to the RXS scans; and ${I}_{\mathrm{calc}}$ are the theoretical profiles for the various terms in the charge order, calculated from Eq.\,\ref{Scattering_tensor_rotation}. Since the RXS intensities cannot be expressed in physical units, there is one degree of freedom left, namely the overall amplitude of the signal; however, a rescaling of the calculated traces will occur if all the symmetry magnitude parameters (${\delta}_{s}$, ${\delta}_{s'}$, and ${\delta}_{d}$) are multiplied by the same factor. Consistently, the results of our fits are always expressed as ratios of the magnitude terms. This additional degree of freedom explains the pre-factor $N-3$, which comes from the fact that the sum of squares of the residuals (${\chi}^{2}$) is normalized to yield the reduced chi-square ${\chi}^{2}_{\mathrm{red}}$, with normalization factor $N-p-1$, where $N$ is the size of the dataset and $p$ is the number of parameters (in this case $p\!=\!2$, since we only consider combinations of two symmetry terms).

\noindent
With this definiton in hand, we have subsequently calculated the cumulative reduced chi-square ${\chi}^{2}_{\mathrm{tot}}$ for the entire dataset (inclusive of all investigated compounds):
\begin{eqnarray}
{\chi}^{2}_{\mathrm{tot}} &=& {\chi}^{2}_{\mathrm{red}} ({\mathrm{Bi2201}}) + {\chi}^{2}_{\mathrm{red}} ({\mathrm{Y651}}) + {\chi}^{2}_{\mathrm{red}} ({\mathrm{Y675}})
\end{eqnarray}

The values for the cumulative reduced ${\chi}^{2}_{\mathrm{tot}}$ (Table\,\ref{Chi_square_analysis}) are used to extract the probability $P$ that the models considered yield a better agreement than a dataset randomly generated from a normal distribution (with mean-square deviations ${\sigma}_{p}^{2}$). These probability levels $P$ can be evaluated based on the cumulative distribution function for a ${\chi}^{2}_{\mathrm{red}}$-distribution. The values for $P$ (Table\,\ref{Cumulative_probability}) are the same also reported in Table\,I in the main text. The outcome for Bi2201 using our model is not conclusive since it yields very similar probability levels for the three 2-component combinations of symmetry terms. This might be due to the larger degree of disorder leading to weaker CDW features in RXS and thus to additional noise and scatter in the experimental data for the azimuthal dependence of the RXS intensities. This, in turn, might hinder our capability of resolving the asymmetry in the azimuthal modulation of the RXS intensity, an aspect which instead emerges more clearly in YBCO. However, we note that a complementary approach based on real-space imaging using STM has been successful in detecting \textit{d}-wave bond order in bilayer Bi${}_{2}$Sr${}_{2}$CaCu${}_{2}$O${}_{6+\delta}$, suggesting that a dominant \textit{d}-wave component characterizes the symmetry of the charge order in both YBCO and Bi-based cuprates [15].
\begin{table}[h!]
\begin{center}
\begin{tabular*}{0.8\linewidth}{@{\extracolsep{\fill}}P{0.2\linewidth}P{0.2\linewidth}P{0.2\linewidth}P{0.2\linewidth}@{}}
\hline
\hline
& & & \\
 & \multicolumn{3}{c}{Probability levels $P$ (\%)}\\
& & & \\
 & $ s + {s}^{\prime} $ & $ s + d $ & $ {s}^{\prime} + d $ \\
\hline
Bi2201 & 51 (${s}^{\prime} / s \!=\! 0.19$) & 50 ($s / d \!=\! 0.14$) & 50 (${s}^{\prime} / d \!=\! 0.17$) \\
Y651 & 82 (${s}^{\prime} / s \!=\! 0.01$) & 97 ($s / d \!=\! 0.21$) & 97 (${s}^{\prime} / d \!=\! 0.27$) \\
Y675 & 0.004 (${s}^{\prime} / s \!=\! -0.01$) & 39 ($s / d \!=\! 0.22$) & 41 (${s}^{\prime} / d \!=\! 0.27$) \\
\hline
\textbf{Cumulative} & \textbf{9.7} & \textbf{80.8} & \textbf{82.2} \\
\hline
\textbf{Cumulative (YBCO)} & \textbf{5.6} & \textbf{83.8} & \textbf{85.5} \\
\hline
\hline
\end{tabular*}
\label{table_report}
\end{center}
\caption{{\bf Statistical comparison of CDW models}. Probability levels $P$ for the hypothesis that the models considered fit the experimental data better than a random sample. The ratios of symmetry components are reported in brackets. The values suggest that a combination of \textit{d}-wave bond-order with either \textit{s}- or \textit{s'}-wave are associated with a large likelihood of describing the experimental data.}
\label{Cumulative_probability}
\end{table}
\clearpage

\subsection*{Supplementary References and Notes}




\noindent
[1] P. Abbamonte, \textit{et al.}, \textit{Phys. Rev. B} \textbf{74}, 195113 (2006).

\noindent
[2] A. J. Achkar, \textit{et al.}, arXiv:1409.6787 (2014).

\noindent
[3] P. D. C. King \textit{et al.}, \textit{Phys. Rev. Lett.} \textbf{106}, 127005 (2011).

\noindent
[4] J. Rosen, R. Comin \textit{et al.}, \textit{Nat. Comm.} \textbf{4}, 1977 (2013).

\noindent
[5] I. Zeljkovic \textit{et al.}, \textit{Nature Materials} \textbf{114}, 585 (2012).

\noindent
[6] D. G. Hawthorn, \textit{et al.}, \textit{Phys. Rev. B} \textbf{84}, 075125 (2011).

\noindent
[7] A. J. Achkar, \textit{et al.}, \textit{Phys. Rev. Lett.} \textbf{109}, 167001 (2012).

\noindent
[8] A. J. Achkar, \textit{et al.}, \textit{Phys. Rev. Lett.} \textbf{110}, 017001 (2013).

\noindent
[9] K. B. Efetov, H. Meier, C. P\'{e}pin,\textit{ Nat. Phys.} \textbf{9}, 442 (2013).

\noindent
[10] S. Sachdev and R. La Placa, \textit{Phys. Rev. Lett.} \textbf{111}, 027202 (2013).


\noindent
[11] P. Abbamonte, \textit{et al.}, \textit{Nat. Phys.} \textbf{1}, 155 (2005).

\noindent
[12] T. Wu, \textit{et al.}, \textit{Nature} \textbf{477}, 191 (2011).

\noindent
[13] A. J. Achkar, \textit{et al.}, \textit{Phys. Rev. B} \textbf{83}, 081106(R) (2011).

\noindent
[14] J. R. Taylor, An introduction to error analysis, \textit{University Science Books} Chap.\,12 (1997).

\noindent
[15] K. Fujita, \textit{et al.}, \textit{Proc. Natl. Acad. Sci.} \textbf{111}, E3026 (2014).

\end{document}